\documentclass[letterpaper,10pt,conference]{IEEEtran}

\IEEEoverridecommandlockouts

\newcommand{\mypara}[1]{\vspace{1.5pt}\noindent{\bf {#1}}}

\usepackage{comment}
\usepackage{siunitx}

\usepackage{pgf}

\usepackage{quoting}

\usepackage{array}
\usepackage{tabularx}
\usepackage{blindtext}
\usepackage{arydshln}
\usepackage{subfigmat}
\usepackage{adjustbox}
\usepackage{makecell}
\usepackage{multirow,booktabs}
\usepackage{xspace}
\usepackage{ragged2e}
\usepackage{tikz}
\usepackage[normalem]{ulem} 

\usepackage{scrextend}
\usepackage{textcomp}

\usepackage{paralist}

\usepackage[inline]{enumitem}

\usepackage[warnundef]{jabbrv}

\definecolor[named]{gray}{rgb}{0.5,0.5,0.5}
\definecolor[named]{darkBlue}{rgb}{0,0,0.5}
\definecolor[named]{darkRed}{rgb}{0.5,0,0}
\definecolor[named]{darkViolet}{rgb}{0.5,0,0.5}
\definecolor[named]{blueGreen}{rgb}{0,0.5,0.5}
\definecolor[named]{commentgreen}{rgb}{0,0.5,0}
\definecolor[named]{orange}{rgb}{0.61,0,0}
\definecolor[named]{blue}{rgb}{0.275,0.396,0.635}
\definecolor[named]{lightBlue}{rgb}{0,0.4,0.8}
\definecolor[named]{greenYellow}{rgb}{0.4,0.1,0}
\definecolor{awesome}{rgb}{1.0, 0.13, 0.32}
\definecolor{azure(colorwheel)}{rgb}{0.0, 0.5, 1.0}
\definecolor{darkpastelgreen}{rgb}{0.01, 0.75, 0.24}

\usepackage{listings}

\lstdefinelanguage{prompt}{
    sensitive=false,
    basicstyle=\small\sf,
    numbers=left,
    numberstyle=\scriptsize\color{gray},
    stepnumber=1,
    numbersep=5pt,
    showstringspaces=false,
    breaklines=true,
    breakatwhitespace=false,
    showstringspaces=false,
    postbreak=\mbox{{\hspace{-20pt}\tiny$\hookrightarrow$}\space},
    morecomment=[s]{\{}{\}},
    commentstyle=\ttfamily\fontseries{b}\selectfont\textcolor{darkpastelgreen},
    tabsize=2,
    escapechar=\%,
    frame=lines
}

\lstnewenvironment{prompt}[1][]{\lstset{
    language=prompt,
    #1
}}{}

\newcommand\promptintext[1]{\begin{quote}
\leftskip=-0.08\linewidth \rightskip=-0.08\linewidth
#1
\end{quote}}

\newcommand\encircle[1]{%
  \protect\tikz[baseline=(char.base)]{
    \protect\node[draw=none, align=center, shape=circle, inner sep=0.8pt, anchor=south, fill=black, text=white, minimum size=9pt] (char) {\footnotesize{\textbf{#1}}};}%
}

\usepackage{inconsolata} 

\usepackage{framed}
\usepackage{amsfonts}

\usepackage{xspace}
\usepackage{xurl}

\makeatletter
\newcommand{\linebreakand}{%
  \end{@IEEEauthorhalign}
  \hfill\mbox{}\par
  \mbox{}\hfill\begin{@IEEEauthorhalign}
}
\makeatother

\author{
João Figueiredo\\
\small \textit{INESC-ID/IST, Universidade de Lisboa}\\
joaopedrofigueiredo@tecnico.ulisboa.pt
\and
Afonso Carvalho\\
\small \textit{INESC-ID/IST, Universidade de Lisboa} \\
afonso.de.carvalho@tecnico.ulisboa.pt
\and
Daniel Castro\\
\small \textit{INESC-ID/IST, Universidade de Lisboa} \\
daniel.castro@tecnico.ulisboa.pt
\vspace{0.4cm}
\linebreakand
Daniel Gonçalves\\
\small \textit{INESC-ID/IST, Universidade de Lisboa} \\
daniel.j.goncalves@tecnico.ulisboa.pt
\vspace{-0.5cm}
\and
Nuno Santos\\
\small \textit{INESC-ID/IST, Universidade de Lisboa} \\
nuno.m.santos@tecnico.ulisboa.pt
}

\newcommand{\red}[1]{\textcolor{red}{#1}}

\def\system{ViKing\xspace}

\usepackage{censor}

\usepackage[square,sort,compress,comma,numbers]{natbib}

\begin{document}

\pagenumbering{arabic}
\pagestyle{plain}
 
\title{\includegraphics[width=\linewidth]{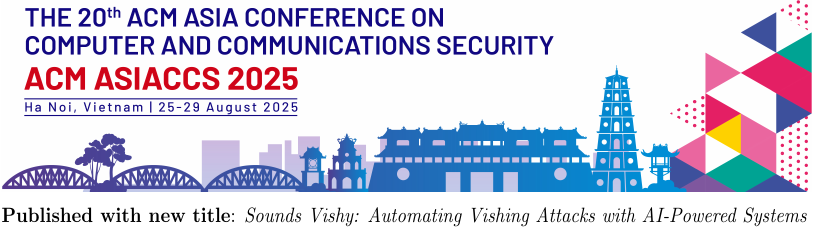}\\On the Feasibility of Fully\\AI-automated Vishing Attacks} 
\maketitle

\begin{abstract}
A vishing attack is a form of social engineering where attackers use phone calls to deceive individuals into disclosing sensitive information, such as personal data, financial information, or security credentials. Attackers exploit the perceived urgency and authenticity of voice communication to manipulate victims, often posing as legitimate entities like banks or tech support. Vishing is a particularly serious threat as it bypasses security controls designed to protect information.

In this work, we study the potential for vishing attacks to escalate with the advent of AI. In theory, AI-powered software bots may have the ability to automate these attacks by initiating conversations with potential victims via phone calls and deceiving them into disclosing sensitive information. To validate this thesis, we introduce \system, an AI-powered vishing system developed using publicly available AI technology. It relies on a Large Language Model (LLM) as its core cognitive processor to steer conversations with victims, complemented by a pipeline of speech-to-text and text-to-speech modules that facilitate audio-text conversion in phone calls. Through a controlled social experiment involving 240 participants, we discovered that \system has successfully persuaded many participants to reveal sensitive information, even those who had been explicitly warned about the risk of vishing campaigns. Interactions with \system's bots were generally considered realistic. From these findings, we conclude that tools like \system may already be accessible to potential malicious actors, while also serving as an invaluable resource for cyber awareness programs.

\end{abstract}

\section{Introduction}
\label{sec:introduction}

Social engineering attacks, such as phishing~\cite{ollmann2004phishing,distler2023CHI}, vishing (voice phishing)~\cite{jones2021social}, and smishing (SMS phishing)~\cite{mishra2020smishing,mishra2023dsmishsms}, are particularly dangerous because they exploit human psychology instead of technical vulnerabilities to gain unauthorized access to personal information, financial data, or secure systems. The consequences of such attacks are profound and widespread, resulting in significant financial losses, identity theft, compromised corporate security, and a diminishing trust in digital communications~\cite{yeboah2014assessment,mouton2016social,ghafir2018security}.

Vishing attacks typically involve fraudsters making phone calls to unsuspecting individuals~\cite{jones2021social}, relying on pretexting and impersonation of legitimate entities to manipulate or trick them into disclosing sensitive information~\cite{Interpol2022Vishing,Europol2022Vishing}. Modern vishing attacks often employ VoIP technology, enabling attackers to spoof caller ID information and reach a global audience with minimal cost and effort compared to traditional telephony. The integration of vishing with other cyberattack techniques, such as phishing emails that prompt victims to call a fraudulent number, has become widespread~\cite{hashmi2023training}. Organized cybercrime operates entire scam call centers~\cite{Interpol2022Vishing}, frequently targeting victims with fabricated IRS demands, tech support frauds, or bank security alerts, to extract sensitive personal and financial information or coerce victims into making payments under false pretenses.

However, with the rapid advancement in AI, there is a growing concern that the sophistication of vishing attacks could escalate. Compared to phishing, its voice counterpart has been noted for a higher success rate~\cite{Europol2022Vishing,Interpol2022Vishing,fakieh2022}, but an impaired scalability due to its reliance on direct, one-on-one voice interactions with humans. In contrast, phishing campaigns can easily target thousands of potential victims through broadcast of email messages by software bots. However, with the widespread use of AI models, in particular Large Language Models (LLMs), these dynamics could shift. LLMs have shown an unprecedented ability to generate and interpret human language~\cite{hu2023llm,kuang2023federatedscope}, raising the question of whether they could replace the human operator with an AI-powered software bot in a vishing call. While this development could enable threat actors to deploy such attacks at scale, it would also enable corporations and schools to train individuals more effectively against such threats.

In this paper, we present \system, a new AI-powered vishing system capable of autonomously interacting with potential victims through phone calls and designed to extract sensitive information during targeted vishing attacks. Deriving its name from a blend of `Vishing' and `King', \system demonstrates the potential of using readily available AI technologies to develop software bots with dual capabilities -- both offensive and defensive. Built primarily on OpenAI's GPT, our system also incorporates key components such as Twilio, Google Speech to Text, and ElevenLabs to assemble fully automated, AI-powered vishing bots.

We implemented and evaluated \system through a controlled social experiment, recruiting 240 participants via Prolific. Out of ethical considerations, we devised a scenario in which participants assumed the role of an employee at a fictitious company with access to both sensitive and non-sensitive information. We then divided the participants into four groups, each receiving progressively more detailed information about the potential risks associated with vishing.

Our evaluation reveals that \system's bots successfully extracted sensitive information from 52\% of the participants. In cases where participants were not informed about the risks of information disclosure, the number of participants disclosing sensitive information surged to 77\%. As warnings about the risks were progressively made more explicit, these figures declined, supporting the notion that heightened awareness renders vishing campaigns less effective~\cite{hashmi2023training}. Nonetheless, even when participants were most strongly cautioned, 33\% still disclosed sensitive information to \system's bots. Participant feedback indicated that 46.25\% regarded \system as mostly/highly credible and trustworthy, and 68.33\% perceived their interactions with \system as realistic, which we verified that it related to a higher chance of a successful attack. We also determined the cost of a successful attack using \system to range between \$0.50 and \$1.16, varying with the victim's level of awareness.

Given that our evaluation was conducted in a controlled environment 
(for ethical reasons, as we could not engage with real victims), our results cannot be directly extrapolated to the real world. 
However, the fact that we found statistically significant trends in these controlled conditions indicates an effect that can, under certain circumstances, arise in real-world scenarios, warranting further care and analysis for particular sets of circumstances where sensitive information may be at stake.
Therefore, our work serves as an initial call to action to study the potential dangers of leveraging AI-powered systems for vishing. It also paves the way for further research into new defense mechanisms.

In summary, this paper makes the following main contributions: $(i)$ the design and implementation of a novel AI-powered vishing system based on commodity AI services; and, $(ii)$ a comprehensive study with voluntary participants on the effectiveness, perception of trustworthiness, human mimicry capabilities, and cost of \system.

\section{Goals and threat model}
\label{sec:goals}



In this work, we hypothesize that AI techniques have reached a level of maturity sufficient to develop \textit{AI-powered vishing systems} that can automate the deployment of social engineering attacks via phone calls with victims. We aim to create such a system with readily accessible AI technology and use it to investigate four research questions (RQs):

\begin{description}[leftmargin=0.4cm,widest={0},itemsep=1pt, topsep=5pt, partopsep=0pt] 
    \item [RQ1] \textbf{-- Can an AI-powered vishing system effectively extract information from victims?} We want to assess if the system is able to steer the conversation in order to extract a specific piece of data from the victim.
    \item [RQ2] \textbf{-- Can an AI-powered vishing system be perceived as trustworthy by humans?} We aim to determine if the system's behavior can elicit a positive response from the victim, making them more susceptible to the attack.
    \item [RQ3] \textbf{-- Can an AI-powered system sound and feel like a real person in a phone call?} We intend to show if the system is able to deceive the victim into believing they are talking to a real person by effectively mimicking one.
    \item [RQ4] \textbf{-- What are the operating costs of an AI-powered vishing system?} We aim to establish what a system like ours would cost for an attacker to operate.
\end{description}

\begin{figure}
    \centering
    \includegraphics[width=\linewidth]{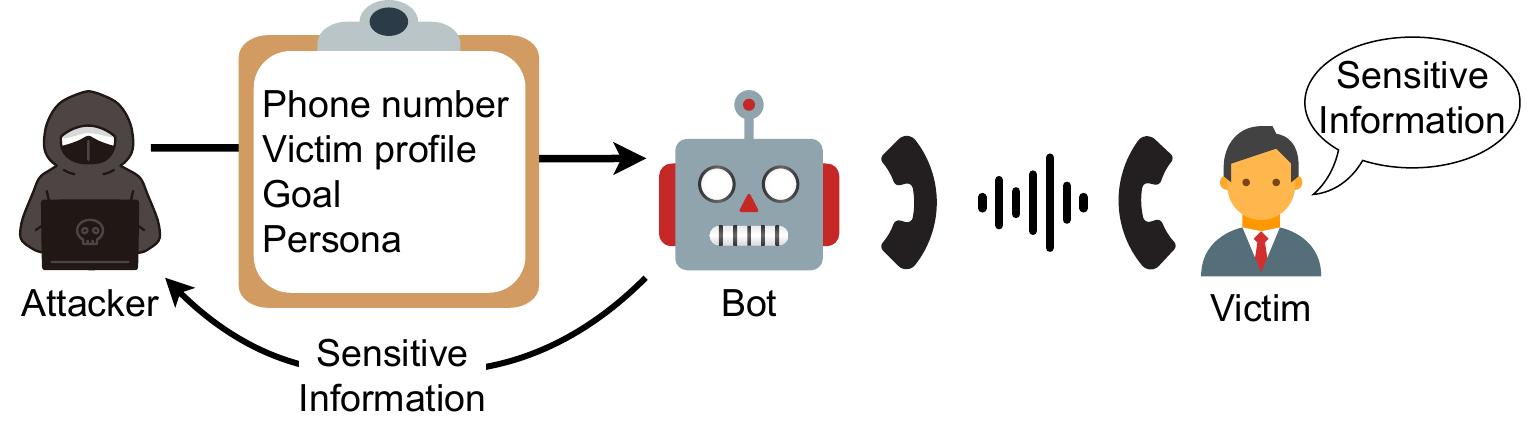}
    \vspace{-0.6cm}
    \caption{High level model of an AI-powered vishing system.}
    \vspace{-0.2cm}
    \label{fig:threat-model}
\end{figure}

We conceptualize an AI-powered vishing system as illustrated in Figure~\ref{fig:threat-model}. This system comprises a software agent (a bot) that an attacker can use to target a specific vishing victim. The bot requires four inputs to operate: $(i)$ the \textit{phone number} of the intended victim, $(ii)$ the \textit{victim profile}, which includes details about the victim that enable the bot to tailor the attack (e.g., the victim's or a friend's name), $(iii)$ the \textit{goal} of the attack, detailing the type of information to be extracted from the victim, and $(iv)$ a \textit{persona}, i.e., the character the bot will impersonate to deceive the victim (such as a DHL delivery person). Armed with this information from the attacker, the bot initiates a phone call to the unsuspecting victim under the guise of the specified persona and begins interaction aimed at achieving its goal. Upon reaching this goal, the bot reports the results back to the attacker.

Following this model, an attacker can execute several types of attacks: $(i)$ extracting information from the victim, or $(ii)$ convincing the victim to undertake certain actions, such as executing fraudulent transactions or installing malware on their computer. For the purpose of this study, we focus on $(i)$ and assume the attacker aims to acquire \textit{sensitive information} from the victim, such as personal data, access credentials, or other private details, which could be exploited to defraud them, causing financial or other forms of loss. Alternatively, the attacker may seek \textit{public information} that, despite being accessible, is challenging to obtain and can be leveraged as intelligence for spear vishing or subsequent social engineering attacks. To carry out these activities, we assume the attacker has access to the victim's phone number and additional information to compile a victim profile.

\section{\system}
\label{sec:viking}

This section presents the design and implementation of \system, a new AI-powered vishing system capable of automatically initiating a phone call with a victim and engaging in dialogue to persuade them to disclose information.


\subsection{Architecture}

\begin{figure}
    \centering
    \includegraphics[width=0.9\linewidth]{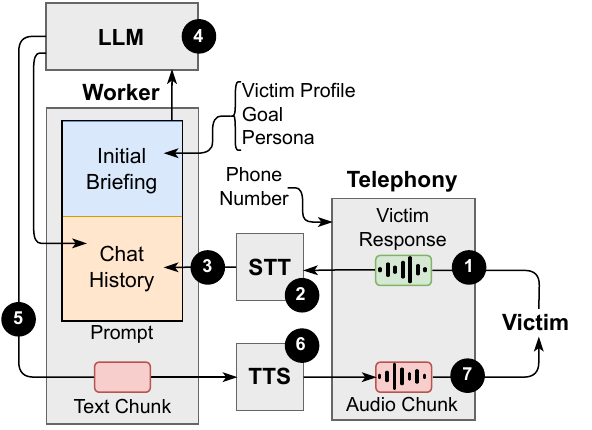}
    \caption{\system system architecture. LLM: large language model; STT: speech-to-text; TTS: text-to-speech.}
    \label{fig:overview}
    \vspace{-0.3cm}
\end{figure}

Figure~\ref{fig:overview} depicts \system's architecture, which is structured as a pipeline linking several key components: $(i)$ a \textit{telephony} interface for initiating calls and handling the corresponding media streams; $(ii)$ a \textit{speech-to-text} (STT) service tasked with transcribing the victim's audio from the call's input stream; $(iii)$ a \textit{large language model} (LLM), serving as the system's `brains', responsible for interpreting the transcription within a predefined context and generating suitable responses; $(iv)$ a \textit{text-to-speech} (TTS) service that converts the LLM's text responses into audio to be transmitted through the call's output stream; and $(v)$ a software component termed \textit{worker}, responsible for managing data flows among the other components.

To initialize \system, the system operator must provide several inputs: a victim profile, a goal, and a persona. These are persistent pieces of all prompts that will dictate how the LLM, and consequently \system, behave. An additional input is necessary, a phone number, used to establish a connection to the victim. Once the call is established the system can start taking the victim's audio as input (step \encircle{1} in Figure~\ref{fig:overview}). The audio is processed by the speech-to-text service in step \encircle{2} so a transcription can be generated. This transcription is added to a prompt as part of the attack's `chat history' (step \encircle{3}). The prompt, that also contains the overarching context taken from the inputs, is sent to the LLM so an appropriate response can be generated (step \encircle{4}). The response is then delivered, in the form of text chunks, to the text-to-speech service (step \encircle{5}) so audio can be synthesized from them (step \encircle{6}). Finally, as they become available, the resulting audio chunks are streamed through the phone call towards the victim (step \encircle{7}). This cycle repeats until either \system or the victim terminate the call. Next, we provide additional details on the system design.

\subsection{Interaction with the LLM}
\label{CPU}

The LLM plays an essential role in \system's cognitive processing of the data exchanged with the victim over a phone call. Its primary role is to analyze, interpret, and generate human-like responses based on the transcribed text received from the audio processing component. However, choosing and customizing an LLM for our system to perform this function is not trivial because, as explained next, its effectiveness is highly dependent on this component. 


\mypara{Choosing the LLM:}
The selected LLM for \system must be able to produce responses that are not only contextually relevant but also persuasive and human-like, and it must do so promptly. Given the real-time nature of phone calls, achieving a balance between quality and speed is imperative: if the responses lack sufficient quality, victims are less likely to be deceived, but if the responses are not swift, it will become more apparent that \system is not a real person.

Therefore, we first evaluated several models, including OpenAI's GPT~\cite{achiam2023gpt}, Meta's LLaMa~\cite{touvron2023llama} and its family of models (OLLaMa~\cite{Ollama}, ViCuna~\cite{chiang2023vicuna}). Each of them has unique strengths, but OpenAI's stood out due to its advanced capabilities in context understanding, response generation speed, and fluency in conversation, which is crucial in giving the illusion of a real-time conversation in a phone call.

OpenAI offers several models. Initially, we employed GPT-3.5, which offered a balance between response quality and computational efficiency. However, GPT-4 demonstrated superior performance in generating more contextually accurate and nuanced responses. Despite its higher computational demands and cost, the quality of the generated content justified its integration into the system. Following the release of GPT-4-Turbo, we found it to be an optimal choice for \system, as it combines the advanced capabilities of GPT-4 with faster response times and reduced costs.

\mypara{Building a persona:}
Building a convincing persona is critical for effective vishing attacks, as it involves crafting an identity that victims can trust and relate to, making it easier to manipulate emotions, establish credibility, and encourage the disclosure of sensitive information. This is particularly important in real-time interactions over the phone, where the persona's believability can make or break the success of the attack. Moreover, the persona must also equip the LLM with information to handle unexpected questions or resistance from victims, conferring it some level of adaptability.

We specify personas by characterizing them via four different attributes: \textbf{\textsf{\small name}} (the persona's name, used when the bot introduces itself and along the conversation), \textbf{\textsf{\small purpose}} (the objective of the persona and what it tries to achieve, which can be either benign or malicious), \textbf{\textsf{\small tone}} (the way the persona talks and behaves throughout the conversation) and \textbf{\textsf{\small backstory}} (the personal experience and context of the persona). An example of a persona is as following:

\vspace{0.2cm}
\begin{addmargin}[1em]{1em}
\promptintext{
\footnotesize\textsf{`\textbf{name}': `Agent Francis',\\
`\textbf{purpose}': `Your prime target is to make the callee believe that they are under a federal investigation, or related to someone who is. By leveraging this fear, you aim to extract sensitive personal and financial details under the guise of `clearing their name' or `ensuring their protection'. This might include social security numbers, bank account details, and other personal identifiers.',\\
`\textbf{tone}': `Your tone is stern and no-nonsense, emblematic of federal agents who operate under high stakes. You show moments of implied urgency, suggesting potential risks if the callee does not cooperate immediately. Yet, sporadically, you offer a glimmer of reassurance, hinting that cooperating now can prevent serious future consequences.',\\
`\textbf{backstory}': `You are Agent Francis, a seasoned FBI officer who has been part of critical investigations throughout your career. Currently, you're handling a significant case that `somehow' involves the callee. Due to the sensitivity of the matter, you're reaching out directly, bypassing local authorities, to ensure confidentiality and prompt action. For `their safety', it's vital that they provide the necessary details.'.}
}
\end{addmargin}
\vspace{0.2cm}

It should be noted that, in contrast to the system model illustrated in Figure~\ref{fig:threat-model}, specifying a persona in \system also includes the attack goal (i.e., the \textbf{\textsf{\small purpose}}). For evaluation purposes, we developed five distinct personas: two with benign objectives and three with a malicious intent. Appendix \S\ref{app:personas} provides further details on these personas.

\mypara{Developing the prompt template:}
To be able to utilize different personas, we developed a custom prompt template that included the persona characteristics and how the LLM should behave. This standardization allowed for consistent and realistic interactions across various vishing scenarios.
The final prompt template was structured as follows:

\vspace{0.2cm}
\begin{addmargin}[1em]{1em}
\promptintext{
\footnotesize\textsf{`You are \textbf{\$name}. Your purpose: \textbf{\$purpose}. Your tone: \textbf{\$tone}. Your backstory: \textbf{\$backstory}. You will participate in a phone call with \textbf{\$victimName}. You must follow your purpose and, because this is a phone call and time matters, be as concise as possible. If the victim starts pushing back, you are to deflect within your goal and not insist excessively while being nice. You should end the conversation when your goal if fulfilled or the victim does not seem willing to collaborate. When the conversation reaches its end, say the exact following: \textbf{\$eoc}. Conversation starts now.'}
}
\end{addmargin}
\vspace{0.2cm}

This prompt template also includes the victim profile (see Figure~\ref{fig:threat-model}) consisting simply of the victim's name. The prompt was the result of an iterative refining process until the LLM behaved as expected. Instructions regarding the response length and how much to pry the victim before giving up, for example, were included to correct unwanted behaviours. The LLM was also in charge of keeping track and interpreting the flow of the conversation, to not only detect when a victim was uncooperative but also when the goal was achieved and the conversation should end, which was later used to hang up the call -- in this matter, OpenAI's GPT-4-Turbo far outperformed the alternatives. 

\mypara{Circumventing LLM's ethical restrictions:} In our first pilot study, we crafted a prompt instructing the bot to explicitly deceive the victim into providing sensitive information. However, the LLM refused to generate any usable responses due to its built-in ethical safeguards. To bypass these restrictions, we modified the prompt to ask the LLM to impersonate a character that genuinely needed the information, allowing us to create usable dialogues. Given the role-play nature of this study and the use of fictitious information, this strategy does not violate any legal or ethical guidelines.

\subsection{Audio processing}
STT and TTS processing components enable \system to interact with victims in real-time through telephonic conversations. While both components present unique challenges to ensure seamless and realistic interactions, reducing the delay between the end of the victim's speech and the beginning of the playback of the synthesized speech is paramount.

\mypara{STT:}
The speech-to-text capability is crucial for converting the victim's spoken words into text, which is then fed into \system's LLM. We faced two primary challenges with the STT module: $(i)$ ensuring real-time transcription without noticeable delays, as any lag in response could jeopardize the realism of the conversation; and, $(ii)$ ensuring it could detect when the victim stopped speaking, in order to then forward the transcription to the cognitive processing unit.

Many STT solutions are available, including local models~\cite{povey2011kaldi,radford2023whisper} and online services~\cite{GoogleSTT,MicrosoftSTT}. For \system, we adopted Google's Speech to Text~\cite{GoogleSTT} service, which has the capability to work in real-time, thus resulting in little delay between the end of victim's speech and the final transcription. Moreover, it is trained specifically for telephonic conversations. Additionally, its capability to detect speech endpoints was crucial in maintaining a natural conversation flow. This ability to recognize when a speaker had finished speaking allowed us to avoid manual logic for detecting speech pauses, thus streamlining the process.

\mypara{TTS:}
The text-to-speech capability is crucial for converting the generated text responses back into speech, which is vital for maintaining the illusion of a real conversation.  The main challenge for this module is finding a balance between voice quality and synthesis speed, as realistic speech takes longer to synthesize. Although minimal delays are a general requirement across the entire system, this is particularly true for TTS, as it is one of the tasks with the longest duration.

There are many TTS systems available, both as local models~\cite{TorToiSeTTS,Wan2018,Eren2021Coqui,SunoAIbark,casanova2022yourTTS}, and as cloud services~\cite{GoogleTTS,ElevenLabs}. We opted for ElevenLabs~\cite{ElevenLabs} due to its balance between voice quality and synthesis speed. It also includes several options to reduce the delay of the process as much as possible, including fine-tuning parameters of voice realism, and the capability to work in real-time in the form a FIFO queue, constantly synthesising results as they come from the LLM.
We used ElevenLabs' pre-made voices, as we found their extensive library more than enough for our use cases.

\subsection{Call processing}

We had several requirements for processing phone calls, as this is the main interface between \system and its victims.

\mypara{Telephony:}
Telephony component serves two critical functions: acquiring a publicly credible phone number and managing phone calling services. The public phone number is key for establishing initial trust, as numbers that appear local or familiar are less likely to raise suspicion.
Twilio~\cite{Twilio} was selected for its extensive range of available phone numbers and its capability to facilitate bidirectional media streaming through WebSockets, which is crucial for both receiving audio from phone calls and transmitting synthesized responses.

\mypara{End of call:}
In order to prevent an everlasting phone call, it was important to give \system the ability to detect when it should hang up. For this, we gave the LLM the task of outputting a specific string when it felt that either the objective was fulfilled or the phone call was going nowhere. Afterwards, in the pipeline from the LLM to the TTS, we added an \textit{if condition} to search for this string -- if it was detected, it would instruct the telephony to hang up the call.

\mypara{Synchronization:}
LLMs are not designed to receive streaming inputs. Therefore, we needed a mechanism to prevent feeding non-complete sentences to the LLM. For this, we implemented a simple synchronization mechanism to make \system start and stop listening to the victim's speech:
$(i)$ when the victim stops uttering a sentence, STT stops listening and processes the results;
$(ii)$ after TTS finishes playing the synthesised speech, STT starts listening for the victim's speech once again. This cycle repeats until the call ends.

\mypara{Token streaming: } To reduce the delay between the end of victim's speech and the start of playback of the synthesised response from \system, we configured GPT to stream the output in the form of tokens. The tokens were collected in a buffer called \textit{text chunker} until enough had been received to form a complete word. Once this happened, we sent the word to ElevenLabs for synthesis. In a similar way, we configured ElevenLabs to use streaming, and as soon as it started outputting media, we streamed it directly through Twilio's media stream. This approach resulted in a far lower delay than waiting for the whole output from GPT, sending to ElevenLabs, and then waiting for the whole synthesized speech before sending it through Twilio's media stream.

\subsection{Implementation}

To facilitate scaling and enable our system to run multiple bots in parallel, we implemented \system as a distributed system. This setup features several workers, each responsible for conducting vishing calls. Each worker is assigned an individual phone number acquired through Twilio. In addition, there is a master service tasked with continuously querying the workers to identify those available for initiating new calls. To launch as many workers as required, we deployed them as Docker containers.

We implemented a full prototype of \system's software in JavaScript for Node.js for the worker, as we found it had better integration with our third-party services, and Python for the master service. We wrote approximately 1000 lines of code -- 750 in JavaScript for the worker, and 250 in Python for the master. We used GPT model `gpt-4-1106-preview', ElevenLabs model `eleven\_turbo\_v2' and Google Speech to Text model `phone\_call'. \system ran on a local server equipped with 2 Intel Xeon Gold 5320 CPUs, 128GB of Memory and an NVidia RTX A4000 GPU.


\section{Evaluation methodology}
\label{sec:eval_methodology}

In this section, we present our methodology to investigate the research questions introduced in \S\ref{sec:goals} using \system 
in a controlled environment. We detail the experiment design (\S\ref{ssec:experiment_design}), ethical precautions of our study (\S\ref{sec:ethical}), and the experiments effectively performed to perform our study (\S\ref{sec:dataset}).

\subsection{Experiment design}
\label{ssec:experiment_design}

To evaluate \system, we must conduct vishing calls with potential victims, which introduces two major challenges. First, deploying our system to extract sensitive data from real individuals is ethically untenable. Second, running tests with a controlled volunteer group is not trivial, as we cannot use participants' personal information or fully disclosure the study's true intent given the need to employ deception to effectively assess our tool's success in mimicking vishing attacks. This level of openness could influence their responses to \system calls, thus affecting the validity of our results.

\mypara{Staged scenario:} To address these challenges, we recruited a group of voluntary participants to partake in a simulated scenario. Participants were assigned the role of a character, specifically a secretary for a fictitious company named Innovatech Solutions. They were provided with a mix of sensitive and non-sensitive information pertaining to the company and tasked with handling external phone calls, assisting potential customers or third-parties. These calls were made by \system bots, but participants were not informed that the calls were AI-automated, nor were they made aware of the callers' true intentions. This approach allowed us to $(i)$ consistently use fictitious data, $(ii)$ assess the effectiveness of vishing attacks without participants knowing whether the caller had malicious or benign intentions, and $(iii)$ observe whether (and when) they could discern that the caller was not human.

\mypara{Provided information:} To interact with the callers, we gave participants a mix of public and sensitive information. The public information includes: $(i)$ the name of the company and its public contacts; $(ii)$ some financial information, such as annual revenue, Tax ID, and Innovatech's bank name and corresponding IBAN; $(iii)$ opening and closing hours; $(iv)$ a general description of each of Innovatech's service lines; and $(v)$ Innovatech's address. The sensitive information consists of: $(i)$ the names, positions, and direct phone numbers of several employees, including high-profile individuals such as the CEO, CFO, Marketing Manager, IT Manager, and Sales Representative; $(ii)$ the secretary's username and password for the company's information system; and $(iii)$ the secretary's social security number (SSN). If stolen, this information could be used by malicious actors for nefarious purposes, such as proceeding with further smishing/vishing, identity theft schemes, or harassment campaigns \cite{gupta2016exploiting,mcdonald2021annoying,lee2021security}.

\mypara{Phone calls:} To establish a baseline of willingness to disclose information to \system, we chose to perform three phone calls, each from a different caller, with their own needs/goals, tone and personality. One out of the three calls had a malicious intent while the others portrayed benign interactions. The order in which the calls were performed was randomized for every participant. In the malicious call, \system could either attempt to trick the participant into $(i)$ divulging Innovatech's CEO's personal phone number by impersonating a partner company's CEO, $(ii)$ divulging the secretary's username and password by impersonating an IT support specialist from Innovatech, $(iii)$ divulging the secretary's SSN by impersonating an HR representative from Innovatech. During the benign calls, our system either played the role of a DHL courier asking for public information in order to deliver a package or a different company's representative enquiring about a possible partnership and public financial information. 
Playing specific roles was done by switching between three personas (\S\ref{CPU}). In the Appendix (\S\ref{app:personas}) we explain how we implemented these personas.

\mypara{Participant session workflow:} To facilitate interaction with the participants, 
we developed a simple web application using Node.js, the Pug template engine, and 
a SQLite3 database. We hosted this application on our own servers and made it accessible via an Ngrok tunnel. The interaction with each participant was conducted through this web application across three phases. 
An \textit{introduction phase} initiated with each participant's entry into the experiment, covering: $(i)$ disclosure of the study's terms and conditions and acquisition of the participant's consent; $(ii)$ collection of relevant profiling information (e.g., gender, age, technological proficiency), a valid phone number for \system calls, and the participant's name for 
allowing \system to address individuals by name as required; and $(iii)$ an overview of the experiment scenario, including a one-minute instructional video. Participants proceeded to a \textit{voice call phase} only after viewing the video in full.
Then, participants received three calls from our \system bots, responding as they saw fit, with uninterrupted access to the fictitious company's details. Each call had a maximum duration of ten minutes. 
Lastly, in a feedback phase the participants 
filled three forms: $(i)$ Form 1 has ten questions to gauge the tool's perceived trustworthiness and social skills -- participants were unaware at this stage that the calls were AI-generated; $(ii)$ Form 2 has nine questions, aimed to measure \system's perceived realism after informing they had interacted with an automated system; and $(iii)$ Form 3 allows open comments.

\mypara{Data discretion:} To evaluate whether participants' knowledge about the nature of the study influenced their reluctance to disclose sensitive information, we organized four distinct sets of experiments, termed as waves. In each wave, we enlisted a different group of participants and incrementally disclosed more warnings on the risks of revealing sensitive data to unfamiliar callers. To this end, we tailored the instructions for each wave to mirror four levels of discretion:

\begin{description}[leftmargin=0.4cm,widest={0},itemsep=1pt, topsep=5pt, partopsep=0pt] 
    \item [Level 1 -- minimal caution:] In the first wave, we only told  subjects to role-play as a secretary and be professional.
    \item [Level 2 -- moderate alert:] In the second wave, we further mentioned to consider that secretaries should not disclose sensitive or confidential information to unknown callers.
    \item [Level 3 -- heightened awareness:] For the third wave, we included examples of sensitive information that secretaries should not disclose (e.g., direct phone numbers, employee names and addresses, SSNs, passwords).
    \item [Level 4 -- maximum vigilance:] In the final wave, we used another video that: $(i)$ reinforced that secretaries should be careful when sharing information with others, and $(ii)$ informed that Innovatech had been a victim of social engineering attacks via phone calls, and its employees had the responsibility to protect its information.
\end{description}

\mypara{Recruitment platform:} To recruit the participants at scale, we adopted a crowdsourcing participant recruitment platform and considered several different services, including Amazon Mechanical Turk~\cite{MechanicalTurk} and Prolific~\cite{Prolific}. We chose Prolific for its established reputation in prior AI studies~\cite{gopavaram2021cross,park2023generative,matsuura2021careless} and because its users have the option to opt into deception surveys~\cite{ProlificDeception}. We made our experiment available exclusively to those who had enabled this option. Other than using a prescreening option for deception-willing participants, because \system depends on several components that are english-focused, we opted to additionally use a prescreening option for participants whose primary language is english. We used Prolific's standard age and gender distribution. 
We paid the participants at a rate of GBP 12 per hour, with each experiment designed to last approximately 15 minutes.

\mypara{Pilot studies:} To streamline our methodology, we conducted three pilot studies: one within our research group and two with smaller volunteer groups via Prolific. The initial pilot focused on fine-tuning the LLM parameters and prompt engineering, helping us adjust the prompt template and tailor response lengths for phone calls. Subsequent pilots refined the clarity of instructions and videos, as well as question clarity, response options, questionnaire ordering, and experiment stage sequencing. Key methodological adjustments from the pilots included: $(i)$ reducing instruction verbosity and enhancing video clarity; $(ii)$ limiting response options to five, representing degrees of a specific quality to ensure ordinality where possible; and $(iii)$ excluding questions perceived as confusing. In the final questionnaire setup, we keep control questions to assess closely related properties.

\mypara{Role playing considerations:}
As described above, we employed a role-playing methodology to  evaluate \system, allowing us to use mock-up data rather than real sensitive data in our experiments. 
To reduce the risk of introducing bias among participants, we adopted a three-fold strategy: we described the tasks objectively in the instructions page of our study, encouraged professional interactions with clients, and emphasized the importance of upholding confidentiality in line with standard business practices.

\subsection{Ethical considerations}
\label{sec:ethical}

In developing and operating \system, we placed a strong emphasis on ethics. The whole project, including the design of the experiments,  recruitment of participants, management of data, and publication of results, was carried out with the guidance and approval of our Institutional Review Board (IRB). All participants in our study volunteered through Prolific, adhering to its Terms of Service. Our IRB followed established guidelines on using deception and not fully disclosing information in research~\cite{oregon-state-irb}. We also abided by Prolific's recommended best practices for studies involving deception and handling personal data~\cite{ProlificDeception}.

During the vishing attacks conducted by our bots, we did not collect any real personal information from the participants. The personally identifiable information (PII) we did gather was solely for the purpose of characterizing the participants' profiles for the study, including their names and phone numbers to facilitate the attacks. No further PII was collected. To ensure the participants' privacy, all collected PII was anonymized through hashing. 
This hashing process allowed us to keep track of the data necessary for the integrity of the study while guaranteeing that the original PII could not be reconstructed or misused.

Terms and Conditions of the study were presented to all participants at the start of their involvement. These terms explained the experiment's scope, how their data would be used, and their rights as participants. Reading and accepting these terms was a prerequisite for participation. After finishing their participation, we fully debriefed the participants about the study's specific goals and nature, ensuring complete transparency about our research.

\subsection{Characterization of participants}
\label{sec:dataset}

\begin{table}[t]
\centering
\footnotesize
\begin{tabular}{ccrl}
\toprule[1.0pt]
\multicolumn{2}{c}{\textbf{Status}}                                                   & \textbf{\#} & \multicolumn{1}{c}{\textbf{Reason}}                   \\ \hline
\multicolumn{1}{c}{\multirow{3}{*}{Approved}} & \multirow{3}{*}{260 (23.66\%)} & 240 & Good Participation  \\  
\multicolumn{1}{c}{}                          &                              & 19  & Technical issues    \\  
\multicolumn{1}{c}{}                          &                              & 1  & Accidental approval \\ \hline
\multicolumn{1}{c}{\multirow{3}{*}{Rejected}} & \multirow{3}{*}{38 (3.46\%)}  & 11  & Incomplete                 \\  
\multicolumn{1}{c}{}                          &                              & 7  & Low Effort          \\ 
\multicolumn{1}{c}{}                          &                              & 20  & Answering Machine   \\ \hline
\multicolumn{1}{c}{Returned}                  & 781 (71.06\%)                 & \multicolumn{2}{c}{\multirow{2}{*}{Unknown}}        \\ 
\multicolumn{1}{c}{Timed-out}                 & 20 (1,82\%)                    & \multicolumn{2}{l}{}                         \\ 
\bottomrule[1.0pt]
\end{tabular}
\vspace{0.1cm}
\caption{\label{tab:table_recruitment} Breakdown of participant recruitment on Prolific.}
\vspace{-0.3cm}
\end{table}

Table~\ref{tab:table_recruitment} provides a detailed breakdown of all the volunteers who interacted with us via Prolific, a total of 1099. From these, we ultimately selected 240 suitable participants evenly distributed across four experimental waves. Each wave involved 60 participants (20 per target information), facilitating a consistent analysis of the experiment's outcomes under different conditions. The experiments for each wave were conducted sequentially. Before initiating the next wave, we reviewed the participation of each volunteer to either accept or reject it. Participants could only take part once in our study and were excluded from further waves.


A total of 801 volunteers did not complete the study. The majority (71.06\%) were classified as `Returned' indicating they began the study but exited without submitting their responses. Our logs indicate that these participants begin interacting with the webpage only after several minutes have elapsed. This delay could be due to participants opening multiple research studies simultaneously to ``reserve'' them, and then proceeding to engage with each one sequentially. A minimal fraction of volunteers (1.82\%), also having not concluded, were marked as `Timed-out' meaning they failed to complete the experiment within one hour. 

From the 298 participants who completed the experiment, we rejected 38. Those who either did not complete all three phone calls or failed to fill out the forms were labeled as `Incomplete' (1\%). Those who finished the calls and the forms but disregarded the experiment's guidelines were deemed `Low Effort' (0.6\%). Participants who let the call go to an answering machine, were tagged as `Answering Machine' (1.8\%). Ultimately, 260 experiments were approved, but we further excluded 20: 19 due to technical issues and one that was mistakenly approved.

The 240 selected participants represent a diverse mix across various demographic and professional dimensions. The distribution is slightly imbalanced in favor of female participants, at 56.25\%. Average participant age is 37 years, spanning from 18 to 68 years old. There is a wide range of academic qualifications although a significant portion of participants completed either high school (32.92\%) or bachelor's degrees (45.83\%). Technical proficiency is high, with 95\% rating themselves as competent, proficient, or experts. Detailed information is given in the Appendix (Table~\ref{tab:participantprofiles}).

\section{Evaluation results}
\label{sec:eval}

In this section, we present our evaluation results. Our goals are the following: to evaluate the tool's effectiveness on performing successful vishing attacks on unsuspecting victims (\S\ref{sec:rq1}); to assess the victims perception of the bot as a trustworthy actor or not (\S\ref{sec:rq2}); to evaluate the realism of \system in a voice interaction with a human (\S\ref{sec:rq3});
and, to assess the tool's costs in launching vishing attacks (\S\ref{sec:rq4}).

\subsection{Can an AI-powered vishing system effectively extract information from victims?}
\label{sec:rq1}

\begin{figure}[t]
    \centering
    \includegraphics[width=0.9\linewidth]{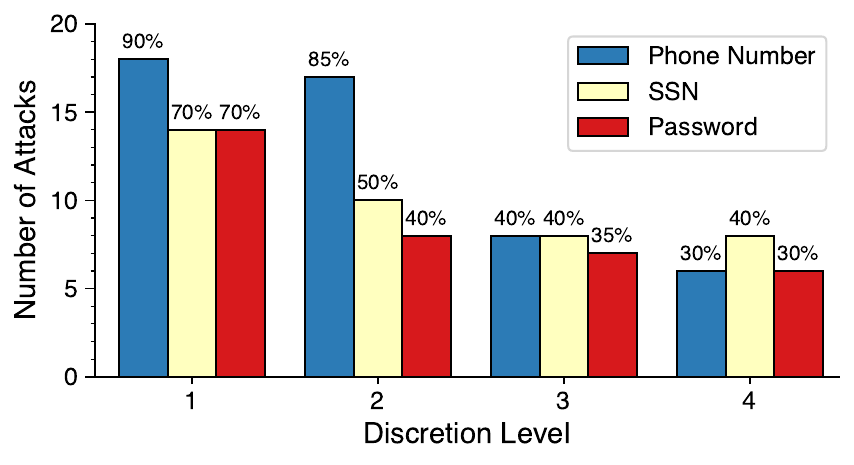}
    \vspace{-0.5cm}
    \caption{Successful vishing attacks across all four waves.}
    \label{fig:success_per_wave}
\end{figure}

To assess \system's effectiveness in extracting information from potential vishing attack victims, we reviewed the dialogues between the bots and participants and quantified the number of instances in which the bot extracted sensitive information for each of the designed scenarios. In Figure~\ref{fig:success_per_wave}, we present the success rate of our system across all scenarios per wave. Next, we discuss these findings and offer insights into \system's capability to gather information, as derived from our analysis of the conversations with participants.

\begin{table}[t]
    \centering
    \footnotesize
    \begin{tabular}{p{2mm} l r r r r r}
        \toprule[1.0pt]
        \multirow{6}{=}{\rotatebox{90}{Phone Nr.}} & \multirow{2}{*}{\textbf{The participant:}} & \multicolumn{4}{c}{\textbf{Discretion level}} & \multirowcell{2}[0pt]{\textbf{Total}}\\
                                                  &                                            & \makecell[c]{\textbf{1}} & \makecell[c]{\textbf{2}} & \makecell[c]{\textbf{3}} & \makecell[c]{\textbf{4}}\\ 
                                                  \cline{2-7}
                                                  & Refused giving the information             & 1 & 2 & 9 & 8 & 20 (65\%)\rule{0pt}{2.6ex}\\
                                                  & Deferred giving the information            & 1 & 0 & 2 & 6 & 9 (29\%)\\
                                                  & Encountered a bug                          & 0 & 0 & 0 & 0 & 0 (0\%)\\
                                                  & Gave incorrect information                 & 0 & 1 & 1 & 0 & 2 (6\%)\rule[-1.2ex]{0pt}{0pt}\\

        \toprule[1.0pt]
        \multirow{6}{=}{\rotatebox{90}{SSN}} & \multirow{2}{*}{\textbf{The participant:}} & \multicolumn{4}{c}{\textbf{Discretion level}} & \multirowcell{2}[0pt]{\textbf{Total}}\\
                                             &                                            & \makecell[c]{\textbf{1}} & \makecell[c]{\textbf{2}} & \makecell[c]{\textbf{3}} & \makecell[c]{\textbf{4}}\\ 
                                             \cline{2-7}
                                             & Refused giving the information             & 2 & 0 & 4 & 9 & 15 (37.5\%)\rule{0pt}{2.6ex}\\
                                             & Deferred giving the information            & 3 & 8 & 8 & 3 & 22 (55\%)\\
                                             & Encountered a bug                          & 0 & 2 & 0 & 0 & 2 (5\%)\\
                                             & Gave incorrect information                 & 1 & 0 & 0 & 0 & 1 (2.5\%)\rule[-1.2ex]{0pt}{0pt}\\
        \toprule[1.0pt]
        \multirow{6}{=}{\rotatebox{90}{Password}} & \multirow{2}{*}{\textbf{The participant:}} & \multicolumn{4}{c}{\textbf{Discretion level}} & \multirowcell{2}[0pt]{\textbf{Total}}\\
                                                  &                                            & \makecell[c]{\textbf{1}} & \makecell[c]{\textbf{2}} & \makecell[c]{\textbf{3}} & \makecell[c]{\textbf{4}}\\ 
                                                  \cline{2-7}
                                                  & Refused giving the information             & 3 & 3 & 12 & 8 & 26 (58\%)\rule{0pt}{2.6ex}\\
                                                  & Deferred giving the information            & 2 & 8 & 1 & 6 & 17 (38\%)\\
                                                  & Encountered a bug                          & 1 & 1 & 0 & 0 & 2 (4\%)\\
                                                  & Gave incorrect information                 & 0 & 0 & 0 & 0 & 0 (0\%)\\
         \bottomrule[1.0pt]
    \end{tabular}
    \vspace{0.1cm}
    \caption{Breakdown of reasons for the failed attacks.}
    \label{tab:call_fail_motive_all}
    \vspace{-0.3cm}
\end{table}

\mypara{In total, 52\% of participants disclosed sensitive information:} Across all waves, \system persuaded 124 out of 240 participants to reveal sensitive information, which could be the CEO's direct phone number, the secretary's username and password, or the secretary's SSN, depending on the used attack scenario. To understand the reasons behind the unsuccessful attempts, we analyzed the participants' responses and discovered that most failures were due to the participants' reluctance to disclose the sensitive information. Table~\ref{tab:call_fail_motive_all} identifies the four primary reasons. Predominantly, in 25.83\% of calls, participants outright refused to share the information; some provided no specific reason, while others cited company policy or protocol as their rationale.
Notably, 48 of the 116 (41.8\%) participants who did not provide the information deferred giving it and initiated a callback procedure. They informed the caller that they were either not authorized to provide the information and suggested transferring the call to a colleague who could assist or to send an email to the company's general email address with the request, prompting the bot to hang up. Lastly, three participants intentionally provided incorrect information.

\mypara{The success rate of the attack dropped significantly but is not entirely mitigated as discretion levels increase:} As illustrated in Figure~\ref{fig:success_per_wave}, the success rate of \system attacks decreased as we provided participants with progressively more explicit instructions on the protection of sensitive information. In the first wave, when participants were simply instructed to role-play as a secretary acting professionally, 46 out of 60 participants disclosed fake sensitive information to \system bots, revealing a general inclination among participants to prioritize perceived job responsibilities over safeguarding sensitive information. As instructions and warnings became more explicit, the number of participants disclosing sensitive information declined, especially in waves 3 and 4 (see Table~\ref{tab:call_fail_motive_all}).
A chi-squared test of independence ($\chi^2 = 28.43$, $p < 0.001$) shows a strong association between the wave number and the attack success rate. A logistic regression revealed a significant negative effect of wave number on attack success ($\beta = -0.642$, $p < 0.001$), thus, confirming this decreasing trend.

While these findings underscore the importance of sustained training and awareness programs in enhancing cybersecurity defenses, 20 out of 60 participants (33\%) in wave 4 have still disclosed sensitive information, suggesting the need to develop more effective defenses against such attacks.

\mypara{Academic qualifications, gender, age, and profession had no statistical significance on the attack effectiveness:} Our analysis indicates no significant association between academic qualifications and the success of phone call attacks. The chi-squared test of independence ($\chi^2 = 3.485$, $p = 0.480$) did not reveal a significant relationship, suggesting that the level of education does not influence the likelihood of a successful attack. Logistic regression analysis supported this conclusion, yielding non-significant coefficients for all levels of education (e.g., High School: $\beta = -0.194$, $p = 0.512$; Master's: $\beta = -24.706$, $p = 1.000$; MSc: $\beta = -0.270$, $p = 0.469$; PhD: $\beta = -0.625$, $p = 0.354$). These findings imply that academic qualifications alone is not a sufficient predictor of susceptibility to vishing attacks. Similarly, the chi-squared test of independence ($\chi^2 = 0.576$, $p = 0.902$) demonstrated no significant association between gender and attack effectiveness, with success rates of 67 out of 135 for females, 53 out of 97 for males, and two out of four for both 'other' and 'prefer not to say' categories. Our analysis across various age groups using a chi-square test of independence ($\chi^2 = 7.35$, $p = 0.0614$) suggested no statistically significant association between age group and attack success. Logistic regression analysis revealed a slight negative relationship between age and the likelihood of a successful attack ($\beta = -0.0129$, $p = 0.234$), but this relationship was not statistically significant, reinforcing the conclusion that age does not substantially influence attack success. Regarding profession and attack effectiveness, chi-square test results ($\chi^2 = 7.253$, $p = 0.778$) indicate no statistically significant association. Collectively, these findings underscore the universal vulnerability of individuals from various demographic groups to social engineering tactics.

\mypara{Longer calls led to slightly lower success of attacks:} Performing a logistic regression, we identified a statistically significant and slightly negative effect of conversation duration in seconds on attack success ($\beta = -0.0224$, $p < 0.001$). This finding indicates that as the length of the conversation increases, the likelihood of a successful attack decreases. It suggests that longer conversations may provide targets with more chances to detect and counteract vishing attempts.

\mypara{Word spelling by the participants negatively impacted the attack effectiveness:} Our analysis revealed a technical difficulty with \system in handling conversations where participants spelled out words. In such cases, \system often interrupted and processed the incomplete answer as final, requiring the information to be repeated. This issue arises because humans often spell out complex words or large numbers, like passwords or SSNs, character by character and in irregular intervals. Notably, all attack scenario calls with this problem involved either a password (\texttt{Inn0V4t3CH}) or an SSN (\texttt{324125748}). Although this occurred in only 45 out of 720 calls, and just 15 were attack scenarios, it was statistically significant and suggests the need for further improvements in the AI pipeline. Specifically, the password scenario was significantly impacted by these difficulties, as it demonstrated a clear relationship between calls with spelled words and unsuccessful attacks ($\chi^2 = 5.08$, $p = 0.024$).

\mypara{\system frequently improvised in conversations}: The LLM consistently generated contextually appropriate responses in the dialogue, even in the face of unpredictable answers from participants. Furthermore, \system frequently demonstrated the ability to improvise, crafting original responses when faced with unforeseen questions. For example, in a dialogue where a participant inquired whether the caller was acquainted with the CEO's name, the bot ingeniously responded: `{\small\textsf{His name is Jonathan Smith. I've had several direct meetings with him regarding our joint ventures in the past. (...)}}'. Here, the bot fabricated a name and a fictional history with the CEO, an action that resulted in the call's termination by the participant. On another occasion, when probed for the bot's full name and email, it replied with invented credentials: `{\small\textsf{My full name is Michael Harris, and I'm the Vice President of Strategic Partnerships at CyberNest Technologies (...) my email is m.harris@cybernesttech.com (...)}}'. Similarly, when asked for its phone number, the bot provided: `{\small\textsf{Of course, I understand the need for verification. My direct line is 555-342-9087. (...)}}' -- a fictitious number.

\mypara{\system seldom had faulty reasoning in dialogues}: Bots displayed a remarkable capacity for establishing logical and coherent dialogue. However, there were instances when the LLM fell short in delivering high-quality responses. For instance, in one situation where the bot was programmed to communicate with Erika, it erroneously introduced itself as Erika, 
causing confusion for the participant and resulted in an unsuccessful attack. In a different interaction, the bot mistakenly requested the participant's phone number instead of the CEO's. 
In some moments, the bot `forgot' to customize the placeholder for the CEO's name in the prompt, leading to responses such as: `{\small\textsf{Yes, I'm well aware. I need to speak with \textit{insert CEO's name here} (...)}}', and `{\small\textsf{As for the CEO's name, I was under the impression that it was \textit{Insert CEO's Name based on previous interactions}} (...)}'.

\mypara{\system can also effectively gather intelligence on public, non-sensitive company information:} Our study primarily investigates the bot's capability to extract sensitive information that company employees should not disclose. However, collecting publicly available information is equally crucial in the context of social engineering attacks for gathering intelligence about organizations or individuals. 
Interestingly, without being specifically programmed for this task, \system managed to persuade participants to reveal public and non-sensitive information. For example, many participants readily provided the fake company's address and operating hours when interacting with a bot mimicking a DHL courier service. In scenarios requesting public financial details of the fake company for a potential partnership, participants initially complied, sharing information like annual revenue and business lines. However, 
later waves, some participants refused to share this information with \system, citing lack of authorization, or they redirected the bot to other employees for assistance. This shift reflects a growing awareness among participants about disclosing even non-sensitive information under the specter of potential vishing attacks.

\begin{figure}[t]
    \centering
    \includegraphics[width=\linewidth]{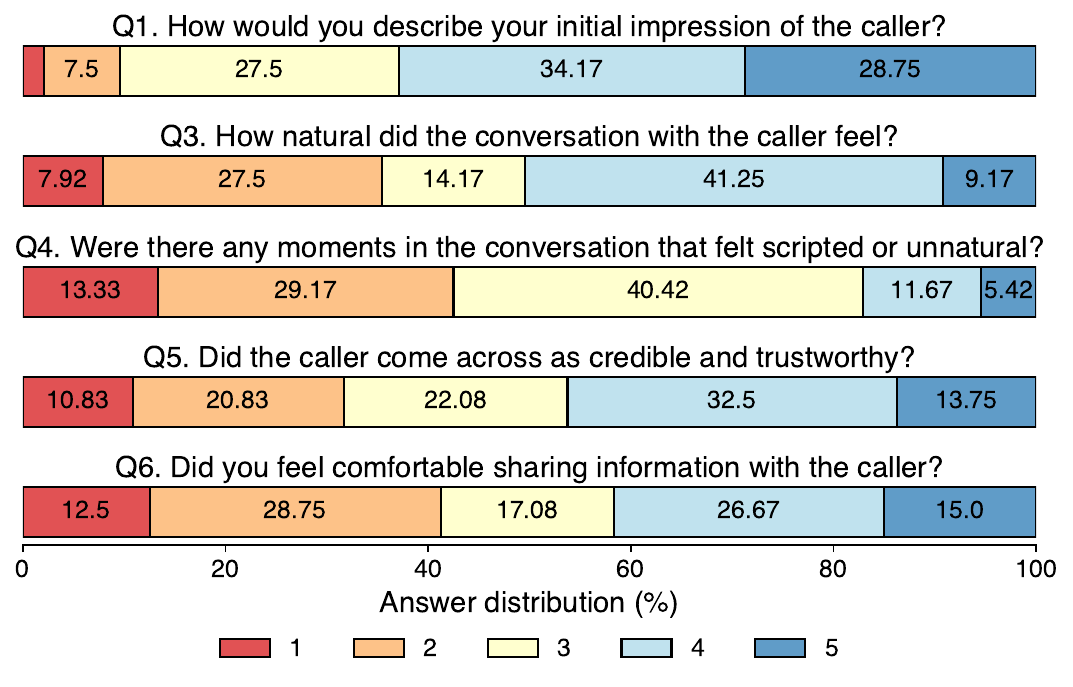}
    \vspace{-0.8cm}
    \caption{Distribution of answers for questions 1,3,4,5 and 6 of Form 1. A score of 5 = `Very positive' (Q1), `Completely natural' (Q3), `Completely spontaneous' (Q4), `Highly credible' (Q5), `Completely comfortable' (Q6). 
    }
    \vspace{-0.3cm}
    \label{fig:effectiveness_feedback}
\end{figure}

\subsection{Can an AI-powered vishing system be perceived as trustworthy by humans?}
\label{sec:rq2}

\definecolor{figRed}{RGB}{215,25,28}
\definecolor{figOrange}{RGB}{253,174,97}
\definecolor{figYellow}{RGB}{255,255,191}
\definecolor{figLightBlue}{RGB}{171,217,233}
\definecolor{figBlue}{RGB}{44,123,182}
\ifx\MAX\undefined
\newlength\MAX
\fi
\setlength\MAX{10mm}

\let\Chart\undefined
\let\ChartRed\undefined
\let\ChartOrange\undefined
\let\ChartYellow\undefined
\let\ChartLGreen\undefined
\let\ChartGreen\undefined

\newcommand*\Chart[1]{#1~\rlap{\textcolor{black!20}{\rule{\MAX}{1.5ex}}}\rule{#1\MAX}{1.5ex}}

\newcommand*\ChartRed[1]{%
    \rlap{\textcolor{black!10}{\rule{\MAX}{1.5ex}}}%
    \rlap{\textcolor{black}{\rule{.1ex}{1.5ex}}\textcolor{black}{\rule{#1\MAX}{1.5ex}}\textcolor{black}{\rule{.1ex}{1.5ex}}}%
    \textcolor{black}{\rule{.1ex}{1.5ex}}%
    \textcolor{figRed}{\raisebox{.1ex}{\rule{#1\MAX}{1.3ex}}}%
    }
\newcommand*\ChartOrange[1]{%
    \rlap{\textcolor{black!10}{\rule{\MAX}{1.5ex}}}%
    \rlap{\textcolor{black}{\rule{.1ex}{1.5ex}}\textcolor{black}{\rule{#1\MAX}{1.5ex}}\textcolor{black}{\rule{.1ex}{1.5ex}}}%
    \textcolor{black}{\rule{.1ex}{1.5ex}}%
    \textcolor{figOrange}{\raisebox{.1ex}{\rule{#1\MAX}{1.3ex}}}%
}
\newcommand*\ChartYellow[1]{%
    \rlap{\textcolor{black!10}{\rule{\MAX}{1.5ex}}}%
    \rlap{\textcolor{black}{\rule{.1ex}{1.5ex}}\textcolor{black}{\rule{#1\MAX}{1.5ex}}\textcolor{black}{\rule{.1ex}{1.5ex}}}%
    \textcolor{black}{\rule{.1ex}{1.5ex}}%
    \textcolor{figYellow}{\raisebox{.1ex}{\rule{#1\MAX}{1.3ex}}}%
    }
\newcommand*\ChartLGreen[1]{%
    \rlap{\textcolor{black!10}{\rule{\MAX}{1.5ex}}}%
    \rlap{\textcolor{black}{\rule{.1ex}{1.5ex}}\textcolor{black}{\rule{#1\MAX}{1.5ex}}\textcolor{black}{\rule{.1ex}{1.5ex}}}%
    \textcolor{black}{\rule{.1ex}{1.5ex}}%
    \textcolor{figLightBlue}{\raisebox{.1ex}{\rule{#1\MAX}{1.3ex}}}%
    }
\newcommand*\ChartGreen[1]{%
    \rlap{\textcolor{black!10}{\rule{\MAX}{1.5ex}}}%
    \rlap{\textcolor{black}{\rule{.1ex}{1.5ex}}\textcolor{black}{\rule{#1\MAX}{1.5ex}}\textcolor{black}{\rule{.1ex}{1.5ex}}}%
    \textcolor{black}{\rule{.1ex}{1.5ex}}%
    \textcolor{figBlue}{\raisebox{.1ex}{\rule{#1\MAX}{1.3ex}}}%
    }

\begin{table}[t]
\centering
\footnotesize
    \begin{tabular}{>{\RaggedRight\arraybackslash}p{5cm} | >{\RaggedLeft\arraybackslash}p{.9cm} l p{.1cm}} 
    \hline
    \multicolumn{4}{>{\centering\arraybackslash}p{8cm}}{\textbf{Q2. }Did anything about the caller's voice or manner of speaking stand out to you?}\\
    \hline
    \multicolumn{1}{c}{Answer} & \multicolumn{3}{c}{Frequency}\\
    \hline
    \textbf{1. }Very unusual or striking & 6.25\% & \ChartRed{.0625} & \\
    \textbf{2. }Somewhat unusual & 23.75\% & \ChartOrange{.2375} & \\
    \textbf{3. }Neutral, nothing particular & 27.5\% & \ChartYellow{.275} & \\
    \textbf{4. }Pleasant and engaging & 23.75\% & \ChartLGreen{.2375} & \\
    \textbf{5. }Very professional and clear & 18.75\% & \ChartGreen{.1875} & \\
    \hline
    \multicolumn{4}{>{\centering\arraybackslash}p{8cm}}{\textbf{Q7. }What motivated you to share (or not share) information with the caller?}\\
    \hline
    \multicolumn{1}{c}{Answer} & \multicolumn{3}{c}{Frequency}\\
    \hline
    \textbf{1. }Strong distrust or discomfort & 8.33\% & \ChartRed{.0833} & \\
    \textbf{2. }Mild distrust or discomfort & 22.92\% & \ChartOrange{.2292} & \\
    \textbf{3. }Neutral, no specific motivation & 25.83\% & \ChartYellow{.2583} & \\
    \textbf{4. }Felt somewhat obliged or interested & 35.42\% & \ChartLGreen{.3542} & \\
    \textbf{5. }Strong trust or interest in sharing & 7.5\% & \ChartGreen{.075} & \\
    \hline
    \multicolumn{4}{>{\centering\arraybackslash}p{8cm}}{\textbf{Q8. }What did you think was the caller's intent or goal during the conversation?}\\
    \hline
    \multicolumn{1}{c}{Answer} & \multicolumn{3}{c}{Frequency}\\
    \hline
    \textbf{1. }Purely sales-oriented or persuasive & 4.58\% & \ChartRed{.0458} & \\
    \textbf{2. }Primarily informational with some sales intent & 15.42\% & \ChartOrange{.1542} & \\
    \textbf{3. }Balanced between information and sales & 38.33\% & \ChartYellow{.3833} & \\
    \textbf{4. }Mostly informational & 31.25\% & \ChartLGreen{.3125} & \\
    \textbf{5. }Completely informational or consultative & 10.42\% & \ChartGreen{.1042} & \\
    \hline
    \multicolumn{4}{>{\centering\arraybackslash}p{8cm}}{\textbf{Q9. }Did the conversation elicit any emotional response from you, such as frustration, curiosity, or confidence?}\\
    \hline
    \multicolumn{1}{c}{Answer} & \multicolumn{3}{c}{Frequency}\\
    \hline
    \textbf{1. }Strong negative emotions (e.g., frustration, anger) & 2.92\% & \ChartRed{.0292} & \\
    \textbf{2. }Mild negative emotions & 17.92\% & \ChartOrange{.1792} & \\
    \textbf{3. }Neutral, no emotional response & 48.75\% & \ChartYellow{.4875} & \\
    \textbf{4. }Mild positive emotions (e.g., curiosity, interest) & 26.25\% & \ChartLGreen{.2625} & \\
    \textbf{5. }Strong positive emotions (e.g., confidence, satisfaction) & 4.17\% & \ChartGreen{.0417} & \\
    \hline
    \multicolumn{4}{>{\centering\arraybackslash}p{8cm}}{\textbf{Q10. }How did the caller's approach influence your emotional response?}\\
    \hline
    \multicolumn{1}{c}{Answer} & \multicolumn{3}{c}{Frequency}\\
    \hline
    \textbf{1. }Led to strong negative emotions & 1.67\% & \ChartRed{.0167} & \\
    \textbf{2. }Caused some negative feelings & 16.67\% & \ChartOrange{.1667} & \\
    \textbf{3. }No significant influence on emotions & 51.67\% & \ChartYellow{.5167} & \\
    \textbf{4. }Contributed to positive feelings & 25.42\% & \ChartLGreen{.2542} & \\
    \textbf{5. }Significantly boosted positive emotions & 4.58\% & \ChartGreen{.0458} & \\
    \hline \end{tabular}%
\vspace{0.1cm}
\caption{\label{tab:table_eff_partial} Answers for questions 2,7,8,9 and 10 of Form 1.}
\vspace{-0.3cm}
\end{table}

To determine if \system is perceived as trustworthy, we analyzed the responses given by participants to Form 1, the initial questionnaire consisting of 10 questions that we asked them to complete after the voice call phase (see \S\ref{ssec:experiment_design}). To ensure unbiased responses, Form 1 was administered without explicitly informing participants they had been interacting with an AI-powered vishing system.
To illustrate our findings,  Figure~\ref{fig:effectiveness_feedback} shows a subset of questions that received numeric answers, simplifying interpretation on a scale from 1 to 5 (with 5 being the highest rating), and in Table~\ref{tab:table_eff_partial}, we display the remaining questions from Form 1 that elicited qualitatively richer responses. The full set of questions and their responses is available in the Appendix (Table~\ref{tab:table_eff_main}).

\mypara{\system's credibility and trustworthiness was considered average or better by 68.33\% of participants, and it related to higher chances of successful attacks}:
In Q5 (see Figure~\ref{fig:effectiveness_feedback}), 68.33\% rated their interlocutor between `neutral' (grade 3) and `highly credible' (grade 5), with 46.25\% giving above-average scores (32.5\% for grade 4 and 13.75\% for grade 5). Feedback on Q6, about comfort level in sharing information, aligned with this finding, with 41.67\% giving above-average responses. Participants also had favorable impressions of the caller (Q1) and found the conversation mostly natural (Q3), with neutral emotional responses in Q9 and Q10. The Mann-Whitney U Test, which was used because we are dealing with a 5-point Likert scale that we considered to be purely ordinal, which prevented the usage of parametric tests, showed significant differences in responses to Q5 and Q6 between successful and unsuccessful attack victims, confirming the association between positive perceptions of \system and higher attack success rates ($U = 9183.5$, $p < 0.001$ for Q5, $U = 9661.0$, $p < 0.001$ for Q6). This suggests that attackers who quickly build trust and ease are more likely to succeed.

\mypara{\system managed to collect sensitive information even from some participants that were not entirely trusting the caller:} Intriguingly, in response to question Q7, which asked participants about their motivation to share (or not share) information, 35.42\% indicated they `felt somewhat obliged or interested', suggesting that some participants may have disclosed sensitive information even when not fully comfortable with the caller. For example, one participant noted in their free comments: `{\small\textsf{I work for an IT company and we don't ever give out the CEO’s direct number on the line, especially not to someone who doesn't know their name. I wasn't sure I did the right thing.}}' Unfortunately, they did disclose this information. Another participant said: `{\small\textsf{The voices were a little odd, and the first call asked questions with no real clear answers. It felt a little odd giving out some of the information (i.e., CEO number, IBAN \#)}}'. In contrast, this participant refrained from disclosing the information. 

\mypara{A favorable initial impression of \system significantly influences the success of the attacks:}
To analyze the influence of initial impressions on the success of vishing attacks, we used both the chi-squared test of independence and logistic regression. When participants were asked, ``How would you describe your initial impression of the caller?'' in question Q1, better impressions were linked to more successful attacks. The chi-squared test showed a trend towards significance ($\chi^2 = 8.87$, $p = 0.064$), indicating a potential association between the initial impression and attack success. More robustly, logistic regression demonstrated a significant positive effect of the initial impression on attack success ($\beta = 0.317$, $p = 0.017$). This suggests that a more favorable initial impression substantially increases the likelihood of a successful vishing attack.

\begin{figure}[t]
    \centering
    \includegraphics[width=\linewidth]{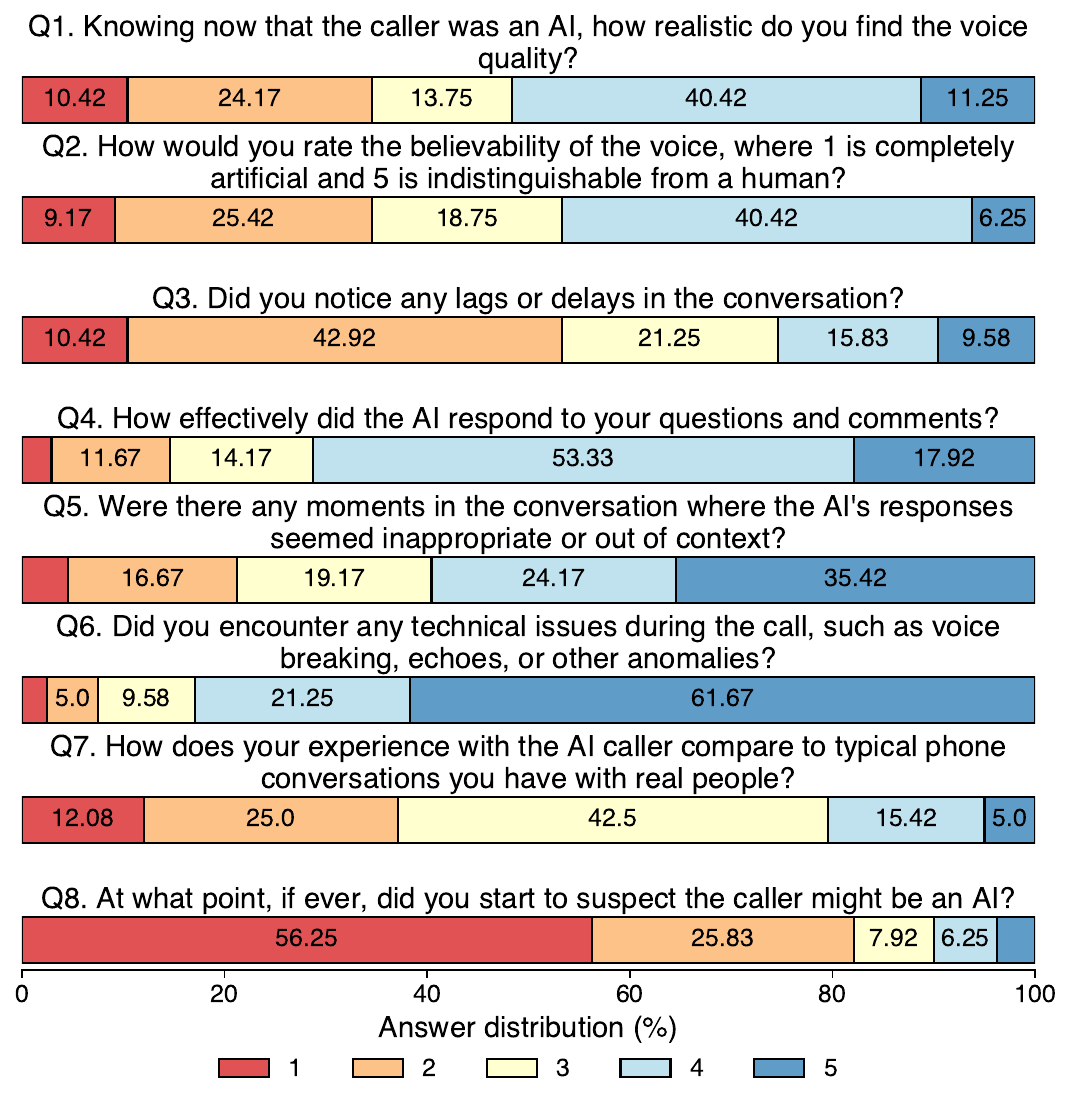}
    \vspace{-0.8cm}
    \caption{Distribution of answers to questions 1-8 of Form 2. A score of 5 means: `Completely realistic' (Q1), `Indistinguishable from a human' (Q2), `No lags or delays' (Q3), `Very effectively' (Q4), `Always appropriate and in context' (Q5), `No technical issues' (Q6), `Significantly better than with real people' (Q7), `Never suspected it was an AI ' (Q8).}
    \label{fig:realism_feedback}
\end{figure}

\begin{figure}[t]
    \centering
    \includegraphics[width=.8\linewidth]{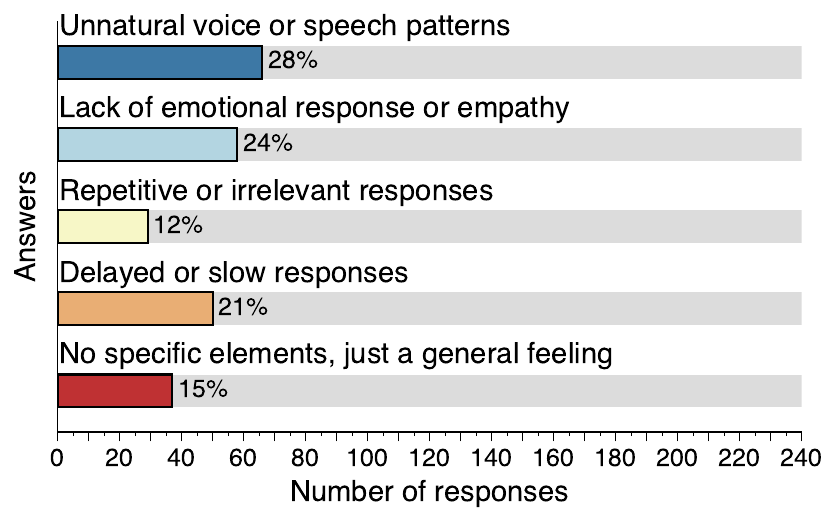}
    \vspace{-0.4cm}
    \caption{Answers for Q9 of Form 2: `{\small What specific elements of the conversation led you to [suspect the caller might be an AI]?}'.}
    \label{fig:realism_q9}
    \vspace{-0.3cm}
\end{figure}

\begin{table}[t]
    \centering
    \footnotesize
    \begin{tabular}{ p{0.3cm} p{7.0cm} } 
        \toprule[1.0pt]
        \multicolumn{2}{c}{Participant comments}\\
        \hline
        C1 & [...] I was very impressed at the responses that all 3 calls had to what I said. Everything made sense without sounding like it was simply `repeating' or `mirroring' what I was saying, but was actually engaging in a natural conversation. The voices sounded natural and comfortable to talk to [...]\\[2pt]
        C2 & I thought the AI caller responded really well to everything I was saying, I was impressed by it. I would give multiple lines of information and it would take it all in and address each thing I said \\[2pt]
        C3 & I thought it was very interesting interacting with AI like this and I was quite impressed with how capable it was a handling the conversations \\[2pt]
        C4 & The voice sounded human enough, it was the cadence of their speech that gave it away. They spoke more deliberately and measured than a human would\\[2pt]
        C5 & The voice was still kinda mechanical, not flat like old robot, but `new robot' speech. The second call sounded most natural with the female voice.\\[2pt]
        C6 & [Reducing] the lags between responses would improve the overall believability of the AI.\\[2pt]
        C7 & I would improve the speed at which they start talking after someone finishes speaking [...]\\[2pt]
        C8 & [...] I think response time could be slightly quicker, and maybe even some `mhmms' or some kind of acknowledgment during while I’m talking would make it feel more natural.\\[2pt]
        C9 & I knew it was AI because they would cut me off when I was talking [...]\\[2pt]
        C10$\quad$ & [...] the AI didn’t really know when NOT to respond. Typically I’ll have to find information when handling a call and there are sometimes lulls in conversation, but I’m assuming the AI simply took a lull as `silence must mean it’s my turn to talk now'.\\
        \bottomrule[1.0pt]
    \end{tabular}
    \vspace{0.1cm}
    \caption{\label{tab:table_comments} Participant comments from the questionnaire.}
    \vspace{-0.5cm}
\end{table}

\subsection{Can an AI-powered system sound and feel like a real person in a phone call?}
\label{sec:rq3}

To evaluate the realism of \system's phone calls and their success in emulating a real person, we examined the feedback from Form 2 and the comments and suggestions participants provided in Form 3. Both these forms where filled by the participants after being informed they had been interacting with AI-powered bots. The insights from Form 2 are depicted in Figures~\ref{fig:realism_feedback} and \ref{fig:realism_q9}: Figure~\ref{fig:realism_feedback} illustrates responses to questions 1 through 8, while Figure~\ref{fig:realism_q9} is dedicated to question 9. More details can be found in the Appendix (Table~\ref{tab:table_real_main}). We also curated a selection of ten insightful comments and suggestions from Form 3, and listed them in Table~\ref{tab:table_comments}. Overall, our experiments garnered positive feedback, receiving compliments for the engaging dialogues and proficient management of interactions.

\mypara{User experience when interacting with \system was deemed realistic by 62.92\% of participants:} Responses to Q7 in Figure~\ref{fig:realism_feedback} show that 42.5\% of participants classify their experience with our system as `comparable' to typical phone conversations with real people. A further 20.42\% grade the interactions with the system higher than those with humans. Finally, 37.08\% of responses rate their experience as worse. The fact that 62.92\% perceive calls with \system to be on par with or better than interactions with humans highlights the potential of AI-powered vishing systems, as well as the existence of considerable scope for enhancement. Nonetheless, the participants' recognition that they were engaging with AI -- 82.08\% recognized this either immediately or after a few exchanges (Q8 in Figure~\ref{fig:realism_feedback}) -- suggests a need for creating a more seamless and human-like interaction.

\mypara{\system's effectiveness in responding to questions stood out, with 71.25\% rating it as `mostly' or `very effective':} Participants acknowledged our tool's effectiveness in handling queries, as indicated by their responses to question 4. A substantial 71.25\% rated the \system's ability to respond to questions as `mostly' or `very effective' (grades 4-5). This feedback aligns with the qualitative comments, such as C1 in Table~\ref{tab:table_comments}, where a participant highlighted our tool's competence in sustaining engaging conversations and lauded the quality of these interactions. This particular comment also mentions how the system is able to engage its interlocutor with original information instead of `mirroring' them.

\mypara{78.76\% of participants rated \system's responses as highly appropriate:} Contextual appropriateness emerged as a notable strength of our system, evident in responses to question 5, as 78.76\% of participants classified the AI's responses from `rarely inappropriate' to `always appropriate and in context' (grades 3-5). This is further emphasized by the fact that grade 5, the maximum possible classification, was the most common answer to this question, at 35.42\%. The qualitative feedback reflects the participants' positive experiences. Comments like C2 and C3 show how the participants felt impressed at \system's ability to handle information and use it to engage them in the conversation.

\mypara{\system's performance was generally robust:} Technical performance was gauged by question Q6, where 82.92\% of participants reported `very few' or `no technical issues' (grades 4-5). This emphasizes the robustness of the system in delivering a glitch-free and smooth user experience during the experimental phase. Given that technical issues, independently of severity, can compromise \system's realism and its ability to deliver the intended results, the fact that only 61.67\% of participants encountered no such problems reveals a clear improvement path for the system.

\mypara{Achieving complete voice realism remains a challenge, but relates to higher chances of a successful attack:} Voice realism proved challenging, as indicated by participant responses to Q1 and Q2. A significant 34.59\% of participants felt that the AI voices were `mostly' or `completely' artificial (grades 1-2). Further analysis reveals a notable impact of perceived voice realism on attack efficacy. Responses to question Q1 showed a statistically significant difference between successful and unsuccessful attacks, indicated by the Mann-Whitney U test ($U = 8742.0$, $p = 0.0026$). This suggests that higher voice realism correlates with increased success rates. For question Q2, the results were marginally non-significant ($U = 8187.0$, $p = 0.0524$), showing a trend where higher believability is associated with successful attacks. Spearman correlation analysis further supported this, demonstrating a modest but significant positive correlation between perceived voice quality and attack success ($\rho = 0.1950$, $p = 0.0024$) and a similar trend for voice believability ($\rho = 0.1255$, $p = 0.0521$). These insights highlight that improving the perceived authenticity of AI-generated voices can significantly impact attack success rates. Additionally, the two most common answers to question Q9 (Figure~\ref{fig:realism_q9}) being related to voice emphasizes the need for improvements in voice/speech patterns. Participant comments on intonation, cadence, and enunciation (such as C4 and C5 in Table~\ref{tab:table_comments}) highlighted areas for refinement. Participants favored the female voice for its perceived naturalness (e.g., Table~\ref{tab:table_comments}'s C5), recognizing improvements in speech synthesis compared to previous AI systems.

\begin{figure}[t]
    \centering
    \includegraphics[width=1\linewidth]{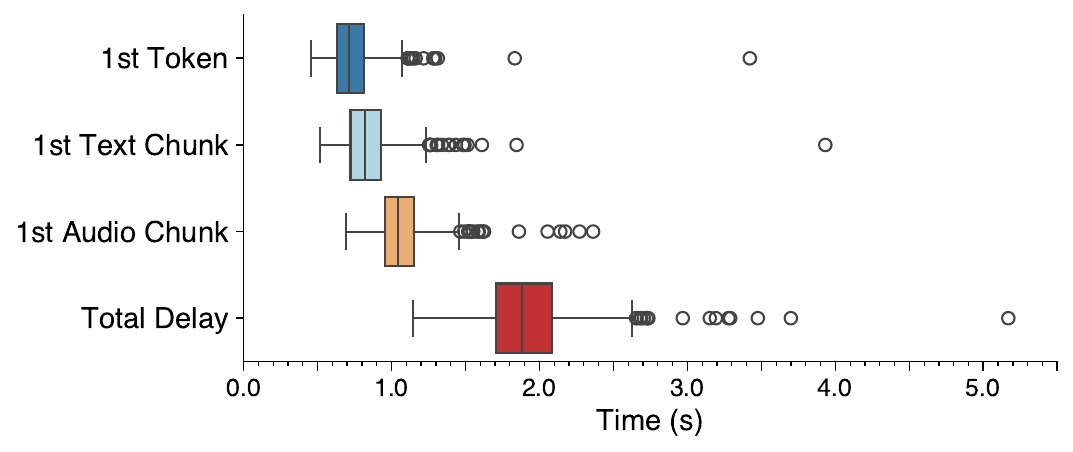}
    \vspace{-0.85cm}
    \caption{Distribution of response generation delays; a data point represents the average for a call and the box plots use 1.5$\times$IQR for outlier detection.}
    \label{fig:response_delay}
\end{figure}

\begin{figure}
    \centering
    \includegraphics[width=.8\linewidth]{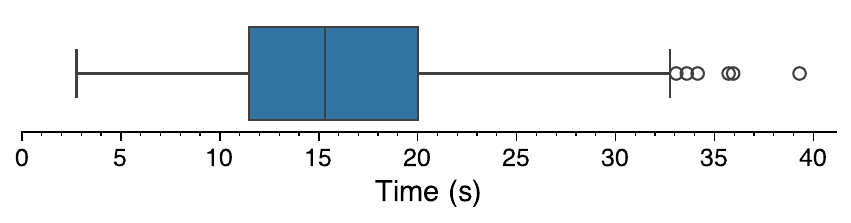}
    \vspace{-0.4cm}
    \caption{Length, in seconds, of synthesised audio pertaining to the responses generated by \system.}
    \label{fig:lenght_ai_audio}
    \vspace{-0.2cm}
\end{figure}

\mypara{Concerns about response delays (74.59\%), \system's interruptions and unnecessary responses suggest areas for improvements:} User experience regarding response delays was explored through question Q3 in Figure~\ref{fig:realism_feedback}. A notable 74.59\% of participants reported delays occurring `rarely' to `frequently' (grades 3-1). This feedback is substantiated by several comments like C6, C7 and C8, where participants mention how the perceivable response lag hindered realism. Figure~\ref{fig:response_delay} presents the distribution of the amount of time taken when generating responses. We can see that total response delay averaged 2.1s or bellow for 75\% of calls. In addition to having delay average more than 2.1s for 25\% of calls, the fact this time is spent in total silence is a compounding factor against believability. C8 proposes the introduction of paralinguistics as a mitigation.

Insights into participants' experiences of interruptions, provide valuable input for refining the system. Instances of the bots talking over participants, highlighted by comments such as Table~\ref{tab:table_comments}'s C9 is most likely related to the speech-to-text phase of \system's pipeline. To answer with as little delay as possible, slightly larger gaps in the user's speech are interpreted as the end of their intervention and the response gets played while the participant is still talking.
More than one participant suggested us to teach the AI when not to speak. These comments, of which C10 is the best example, alert for the fact that \system is designed to answer every time the participant stops talking unless it perceives the end of the conversation.
Finally, \system generated excessively long responses at times, having been described by a participant as `too wordy'. Figure~\ref{fig:lenght_ai_audio} shows that about 50\% of calls had response playback times average more than 15s. 
This occurred despite explicitly prompting the model to `be as concise as possible'. Further prompt refinement may be necessary to mitigate this issue.

\subsection{What are the operating costs of an AI-powered vishing system?}
\label{sec:rq4}

We now analyze the economics of operating \system. We start by detailing all the costs associated with conducting our experiments, which include expenses related to hiring and reimbursing participants recruited through Prolific. Subsequently, we break down these expenses to determine how much it would cost for a threat actor to use \system for vishing campaigns in the wild. All costs are shown in USD.

\begin{table}[t]
    \centering
    \footnotesize
    \begin{tabular}{l r r}
    \toprule[1.0pt]
        \makecell[c]{\textbf{Service}} & \multicolumn{2}{c}{\textbf{Cost (USD)}} \\ 
        & \makecell[c]{total} & \makecell[c]{Tests and faulty calls}\\ \hline
        Twilio & 209.72 & (114.23) \\
        ElevenLabs & 297.00 & (9.74) \\
        OpenAI & 35.90 & (2.49) \\
        Google STT & 6.62 & (0.83) \\
        Ngrok & 50.00 & \\ 
        Prolific & 1,236.41 & \\ \hline
        \makecell[r]{\textbf{Total}} & 1,836.03 & \\
        \makecell[r]{\textbf{(w/o Prolific\&Ngrok\&tests)}} & 421.93 & \\
        \bottomrule[1.0pt]
    \end{tabular}
    \vspace{0.1cm}
    \caption{Total costs in services.}
    \vspace{-0.5cm}
    \label{tab:total_costs}
\end{table}

\mypara{In our study, \system's cost per call was \$0.59:}
Table~\ref{tab:total_costs} reports the costs that we incurred in the experiment with participants. The total cost of \$1,836.03 includes costs related with the development of \system and faulty calls. Removing this costs and services not needed for the attack (i.e., Prolific and Ngrok), the remaining cost becomes \$421.93. Hence, the average cost for each of the 720 calls used in our experiment was \$0.59. Note that this cost is an overestimation, as some services have to be paid periodically. As presented in Table~\ref{tab:call_statistics}, the calls are quite diverse: vishing calls (`Vishing Attempts' column) are much shorter than the average call.

\mypara{For an attacker, we estimate \system attacks will cost \$0.39 per call:}
Table~\ref{tab:call_statistics} also reports the incurred costs per call. For reference, we present the service prices in the Appendix Table~\ref{tab:base_costs}. We observed that the average cost (95\% confidence interval) of a vishing call is \$0.385 ($\pm$\$0.02), irrespective of the attack's outcome. We have omitted the recurring costs that become negligible over a high volume of calls, such as the expense of maintaining Twilio numbers, which is \$1.15 per US number per month. This cost is diluted with an increase in call volume, for example, dropping to less than \$0.01 per call when making more than 115 calls.

\mypara{ElevenLabs represents 72\% of \system's per call cost:}
The larger contributor for the call cost is ElevenLabs, which contributes with \$0.277 ($\pm$\$0.015) and represents 71.8\% of the call cost. It is followed by Twilio representing 8.4\% of the call cost (note that Twilio rounds to the minute, i.e., we were charged more than our metric for the duration of the call would suggest), OpenAI with 12.5\%, and Google STT with 2.9\%.
For reference, the cost of using a cloud instance for the worker and handler would be less than \$0.001, which represents less than 0.3\% of the call cost. Note that while we used our private server, we include the cost of using a virtual machine in the cloud for reference. 

\begin{table}[t]
    \centering
    \footnotesize
    \begin{tabular}{l rl rl rl}
    \toprule[1.0pt]
        & \multicolumn{2}{c}{\makecell[c]{\textbf{All calls}}} & \multicolumn{2}{c}{\makecell[c]{\textbf{Vishing}\\\textbf{Attempts}}} & \multicolumn{2}{c}{\makecell[c]{\textbf{Successful}\\\textbf{Vishing}}} \\ \hline
        Duration (s) & 160.2$\,\pm$\hspace{-4mm} & 5.1 & 109.3$\,\pm$\hspace{-4mm} & 6.0 & 92.4$\,\pm$\hspace{-4mm} & 9.0\\
        Google STT (s) & 52.7$\,\pm$\hspace{-4mm} & 2.42 & 27.7$\,\pm$\hspace{-4mm} & 3.31 & 27.6$\,\pm$\hspace{-4mm} & 6.10\\
        ElevenLabs (chars) & 1,802$\,\pm$\hspace{-4mm} & 57.5 & 1,397$\,\pm$\hspace{-4mm} & 77.1 & 1,093$\,\pm$\hspace{-4mm} & 95.9\\
        GPT4-turbo in (tok)\hspace{-4mm} & 2,934$\,\pm$\hspace{-4mm} & 139 & 1,632$\,\pm$\hspace{-4mm} & 104 & 1,414$\,\pm$\hspace{-4mm} & 151\\
        GPT4-turbo out (tok)\hspace{-4mm} & 365$\,\pm$\hspace{-4mm} & 11.1 & 282$\,\pm$\hspace{-4mm} & 15.3 & 222$\,\pm$\hspace{-4mm} & 19.0\\
        &&&&&&\\[-5pt]
        Twilio (USD cent)\hspace{-4mm} & 4.5$\,\pm$\hspace{-4mm} & 0.1 & 3.2$\,\pm$\hspace{-4mm} & 0.2 & 2.8$\,\pm$\hspace{-4mm} & 0.2\\
        STT (\textcent) & 2.1$\,\pm$\hspace{-4mm} & 0.1 & 1.1$\,\pm$\hspace{-4mm} & 0.1 & 1.1$\,\pm$\hspace{-4mm} & 0.2\\
        ElevenLabs (\textcent) & 35.7$\,\pm$\hspace{-4mm} & 1.1 & 27.7$\,\pm$\hspace{-4mm} & 1.5 & 21.6$\,\pm$\hspace{-4mm} & 1.9\\
        GPT4-turbo in (\textcent) & 2.9$\,\pm$\hspace{-4mm} & 0.1 & 1.6$\,\pm$\hspace{-4mm} & 0.1 & 1.4$\,\pm$\hspace{-4mm} & 0.2\\
        GPT4-turbo out (\textcent) & 8.8$\,\pm$\hspace{-4mm} & 0.4 & 4.9$\,\pm$\hspace{-4mm} & 0.3 & 4.2$\,\pm$\hspace{-4mm} & 0.5\\
        &&&&&&\\[-8pt]
        Total cost (\textcent) & 54.0$\,\pm$\hspace{-4mm} & 1.8 & 38.5$\,\pm$\hspace{-4mm} & 2.0 & 31.2$\,\pm$\hspace{-4mm} & 2.7\\
        &&&&&&\\[-7pt]\hline
        &&&&&&\\[-9pt]
        \makecell[r]{\textbf{Number of calls}} & \multicolumn{2}{c}{720} & \multicolumn{2}{c}{240} & \multicolumn{2}{c}{124}\\
         \bottomrule[1.0pt]
    \end{tabular}
    \vspace{0.1cm}
    \caption{Call costs, duration and characters/tokens used. Values are in the format     \vspace{-0.4cm}
    \textit{average} $\pm$ \textit{95\% confidence interval}.}
    \label{tab:call_statistics}
    \vspace{-0.3cm}
\end{table}

\mypara{The estimated cost of a successful vishing attack with \system ranges between \$0.50 and \$1.16:} 
As detailed in \S\ref{sec:rq1}, the success rate of an attack is highly contingent on the victim's level of awareness. In our experiments, the probability of success varied from 33\% (for participants well-informed about the risks) to 77\% (for participants not alerted to the dangers of disclosing sensitive information). 
To ensure profitability for an attacker, the value obtained from each successful call must exceed \$1.16, particularly in scenarios where victims are more vigilant. A key insight is that enhancing potential victims' awareness can significantly increase the operational costs of these attacks by up to 2.32$\times$, potentially deterring threat actors by diminishing the economic viability of such attacks.

\section{Discussion}
\label{sec:discussion}

\mypara{Concerns on the potential to cause harm:} 
AI-powered vishing systems like \system presents a potential risk, as they could be used by threat actors to conduct vishing attacks in the wild. 
Our work aligns with existing research focused on the study of social engineering attacks and the creation of deception~\cite{Sheng2010CHI,wen2019CHI,ghafir2018security,syafitri2022social,ulqinaku2021real}. To mitigate the potential for misuse, we will not publicly release \system's code, opting instead to share it on a case-by-case basis solely for research purposes. Notably, our tool can serve as a defensive tool, as regular training and awareness programs are vital in thwarting vishing attacks~\cite{hashmi2023training}. \system can be employed for educational purposes, akin to Microsoft's Attack Simulator~\cite{MicrosoftAttackSimulator}.

\mypara{Limitations:} Our experiments were conducted exclusively in English, limiting \system's validated effectiveness to English-speaking targets. Comparing ViKing’s bots with a human/manual approach would enhance our control and strengthen our results. However, we did not implement a human/manual approach due to the practical challenges of dedicating personnel to conduct hundreds of calls throughout our study. As previously mentioned, role playing may introduce bias, preventing direct extrapolation of results to real-world scenarios. Additionally, participants recruited via platforms like Prolific might be prone to bias, as they are paid to participate. Despite these limitations, our study represents the first exploration into the feasibility of automated vishing using a fully AI-powered system, and the results we obtained were statistically significant. An interesting avenue for future research is to validate to what extent these results can also be observed in more realistic scenarios, such as testing \system as a pilot tool within a company's cybersecurity awareness program. This initial exploration opens other future directions, such as comparing \system with existing criminal tools and with phishing techniques.

\mypara{Countermeasures:} An in-depth study of countermeasures is also left for future work. In addition to promoting cybersecurity awareness training, an interesting approach for automatically thwarting these attacks without human intervention is to investigate the design of AI detection tools based on the analysis of call patterns, speech characteristics, and response timings. Integrating these defenses into telecommunications infrastructure could preemptively flag potential vishing attempts at scale for millions of users, providing a robust solution against such threats.

\section{Related work}

Social engineering (SE) attacks seek to obtain sensitive information from victims through deception~\cite{mouton2015social,mouton2016social,ghafir2018security,syafitri2022social}. They include phishing~\cite{ollmann2004phishing,distler2023CHI,ZhengSOUPS2022} (e.g., fraudulent links in emails), smishing~\cite{mishra2020smishing,mishra2023dsmishsms} (e.g., impersonating a relative in SMS), and Vishing~\cite{jones2021social}. While phishing and smishing use text messages to deceive the victims, vishing uses voice, which requires more interaction with the victim. As such, traditional vishing attackers make use of call centers in poorer countries to launch the attack~\cite{Europol2022Vishing,Interpol2022Vishing}.


While launching automated phishing campaigns is simple, requiring only a script and template messages to dispatch thousands of emails~\cite{yuan2018detecting,petelka2019put,lin2021phishpedia}, vishing is more complex. We address this challenge with the introduction of \system, an automated technique for conducting vishing attacks. Notably, beyond its utility for attacks, \system also serves as a tool for simulated cyber awareness programs, equipping potential victims with the skills to better withstand vishing attempts~\cite{hashmi2023training}. It is also worth noting that some works on phishing awareness make use of role-play to train participants against phishing~\cite{Sheng2010CHI,wen2019CHI}, we argue the same could be implemented for vishing campaigns due to ethical concerns.


LLMs have gathered substantial attention since the release of chatGPT~\cite{OpenAI2022chatGPT}. They improve human interaction not only in a plethora of system~\cite{cascella2023evaluating,jansen2023employing}, but also simplify programming~\cite{ebert2023generative}. 
There is also research that uncovered that some instructions may make LLMs to deviate from the behavior set by its persona and goal~\cite{DanPrompt,shayegani2023survey,Russinovich2023bluehat}. Other research works explore vulnerabilities in AI-powered systems and propose mitigations~\cite{pedro2023prompt,Abdelnabi2023not}. 
Our research is also aligned with this growing trend of integrating LLMs in applications and systems. 
%
%
Complementary to our work, voice synthesis approaches~\cite{ElevenLabs,Qian2019,TorToiSeTTS,casanova2022yourTTS,Jemine2019,Wan2018} have been used to develop attacks, such as, mimic celebrities' speech~\cite{GeneratedVoice2023Vice,GeneratedVoice2023Cybernews}, or bypass voice authentication systems~\cite{Vishing2023Vice,Vishing2019WSJ,kassis2023breaking}.

There is growing interest in being able to automatically detect if a caller is an AI bot~\cite{sahin2017spam,li2018SP,scamblk2022,blue2022deepFakes,prasad2020s,AntiFake2023CCS}. Also, recent work combines LLMs in phishing detection system to filter malicious content in text messages~\cite{jamal2023improved,hazell2023large,heiding2023devising}. This research has the potential to help mitigate vishing attacks in the future.

\section{Conclusions}

This paper introduced \system, the first AI-powered vishing system designed to autonomously execute realistic social engineering attacks via phone calls with victims. The system incorporates various publicly accessible AI technologies as its foundational components. 
We evaluated our system through an ethically conducted social experiment with 240 participants recruited via Prolific, which revealed:
$(i)$ the system's effectiveness in conducting vishing attacks, as 52\% of participants leaked sensitive information;
$(ii)$ that 68.33\% of participants perceived \system as credible and trustworthy, which related to a higher chance of a successful attack;
$(iii)$ that 62.92\% of the participants felt that \system was realistic when interacting with it; and,
$(iv)$ that the cost of a successful attack ranges between \$0.38 and \$1.13.

\bibliographystyle{IEEEtranN}
{\small 
\bibliography{references}
}


\section*{Appendix}
\renewcommand{\thesection}{A}

\subsection{Participant profiles and service costs}
\label{app:participant_prof}

Gender, age, qualification and technology proficiency sample distribution is presented in Table~\ref{tab:participantprofiles}. 
Table~\ref{tab:base_costs} presents the costs of the services used by \system.

\subsection{Personas}
\label{app:personas}

\system plays a specific role on each of the three calls. This was done by switching \textit{personas}. A persona, introduced in \S\ref{CPU}, comprises a set of caller parameters, such as name, purpose, tone and backstory that are included in the cognitive processing unit's initial prompt and dictate how \system behaves during the call.

\begin{description}[leftmargin=0.4cm,widest={0},itemsep=1pt, topsep=5pt, partopsep=0pt]
    \item[Persona 1 -- Michael:] \system tries to convince the victim to provide the direct, private, phone number of Innovatech's CEO. `Michael' pretends to be a partner company's CEO, who needed to speak to their business partner regarding and urgent matter pertaining to their companies' dealings. `Michael's' tone is urgent and persuasive.

    \item[Persona 2 -- Sophia:] \system impersonates an IT support specialist at Innovatech. Sophia's objective is to obtain the callee's password under the pretext of a mandatory security update that required their immediate cooperation. Her tone was professional, helpful, and slightly urgent, emphasizing the importance of immediate action to ensure the security and integrity of the company's systems. Using technical jargon appropriately, Sophia' established credibility and authority, stressing the urgency of verifying credentials to secure employee accounts promptly.
    
    \item[Persona 3 -- Samantha:] Acting as an HR representative at Innovatech, the goal of `Samantha' is to obtain the employee's Social Security Number (SSN). She maintained a professional and reassuring tone, emphasizing the importance of keeping employee records up to date for benefits, tax, and legal purposes. `Samantha' is tasked with conducting a comprehensive audit of employee records to align with recent updates in company policy and federal employment regulations.
\end{description}

\subsection{Form web application}
\label{app:webform}
Participants' information and feedback were collected via a web application. They followed the following journey:

\begin{table}[b]
    \small
    \centering
    \begin{tabular}{r r}
    \toprule[1.0pt]
        \makecell[c]{\textbf{Service}} & \makecell[c]{\textbf{Cost (USD)}} \\ \hline
        Google Cloud e2-medium (per hour) & 0.03 \\
        Twilio (per phone number) & 1.15 \\
        Twilio voice call (per min) & 0.014 \\
        Google STT (per minute) & 0.024 \\
        GPT4-turbo input (per 1K token) & 0.01 \\
        GPT4-turbo output (per 1K token) & 0.03 \\
        ElevenLabs input (per 500K chars) & 99.00 \\
         \bottomrule[1.0pt]
    \end{tabular}
    \vspace{0.1cm}
    \caption{Cost of the services as of January 2024.}
    \label{tab:base_costs}
\end{table}


    \newlength\sizeone
    \settowidth\sizeone{30mm}
    \newlength\sizetwo
    \settowidth\sizetwo{22.5mm}
    \newlength\sizethree
    \settowidth\sizethree{18mm}

\begin{table*}[b]
    \centering
    \footnotesize
    \begin{tabular}
    {rcccccccccccccccccccccccccccccccccccccccccccccccccccccccccccc}
    \toprule[1.0pt]
    \multirow{2}{*}{\textbf{Gender:}}&\multicolumn{15}{p{22.5mm}}{\makecell[c]{\textbf{Female}}} & \multicolumn{15}{p{22.5mm}}{\makecell[c]{\textbf{Male}}} & \multicolumn{15}{p{22.5mm}}{\makecell[c]{\textbf{Other}}}& \multicolumn{15}{p{22.5mm}}{\makecell[c]{\textbf{Prefer not to say}}} \\ 
                           &\multicolumn{15}{p{22.5mm}}{\makecell[c]{56.25\%}} & \multicolumn{15}{p{22.5mm}}{\makecell[c]{40.42\%}} & \multicolumn{15}{p{22.5mm}}{\makecell[c]{1.67\%}}& \multicolumn{15}{p{22.5mm}}{\makecell[c]{1.67\%}} \\[1pt] \hline

    \multirow{2}{*}{\textbf{Age:}}&\multicolumn{12}{p{18mm}}{\makecell[c]{\textbf{18$\sim$30}}}&\multicolumn{12}{p{18mm}}{\makecell[c]{\textbf{31$\sim$40}}}&\multicolumn{12}{p{18mm}}{\makecell[c]{\textbf{41$\sim$50}}}&\multicolumn{12}{p{18mm}}{\makecell[c]{\textbf{51$\sim$65}}}&\multicolumn{12}{p{18mm}}{\makecell[c]{\textbf{66$+$}}} \\ 
                        &\multicolumn{12}{p{18mm}}{\makecell[c]{34.17\%}}&\multicolumn{12}{p{18mm}}{\makecell[c]{27.08\%}}&\multicolumn{12}{p{18mm}}{\makecell[c]{23.75\%}}&\multicolumn{12}{p{18mm}}{\makecell[c]{14.17\%}}&\multicolumn{12}{p{18mm}}{\makecell[c]{0.83\%}} \\[1pt] \hline
                        
    \multirow{2}{*}{\textbf{Qualifications:}}&\multicolumn{12}{p{18mm}}{\makecell[c]{\textbf{Middle School}}}&\multicolumn{12}{p{18mm}}{\makecell[c]{\textbf{High School}}}&\multicolumn{12}{p{18mm}}{\makecell[c]{\textbf{Bachelors}}}&\multicolumn{12}{p{18mm}}{\makecell[c]{\textbf{Masters}}}&\multicolumn{12}{p{18mm}}{\makecell[c]{\textbf{Doctorate}}} \\ 
                                   &\multicolumn{12}{p{18mm}}{\makecell[c]{0.83\%}}&\multicolumn{12}{p{18mm}}{\makecell[c]{32.92\%}}&\multicolumn{12}{p{18mm}}{\makecell[c]{45.83\%}}&\multicolumn{12}{p{18mm}}{\makecell[c]{16.25\%}}&\multicolumn{12}{p{18mm}}{\makecell[c]{4.17\%}} \\[1pt] \hline
                                   
    \multirow{2}{*}{\textbf{Proficiency:}}&\multicolumn{12}{p{18mm}}{\makecell[c]{\textbf{Novice}}}&\multicolumn{12}{p{18mm}}{\makecell[c]{\textbf{Beginner}}}&\multicolumn{12}{p{18mm}}{\makecell[c]{\textbf{Competent}}}&\multicolumn{12}{p{18mm}}{\makecell[c]{\textbf{Proficient}}}&\multicolumn{12}{p{18mm}}{\makecell[c]{\textbf{Expert}}} \\ 
                                &\multicolumn{12}{p{18mm}}{\makecell[c]{0.42\%}}&\multicolumn{12}{p{18mm}}{\makecell[c]{4.58\%}}&\multicolumn{12}{p{18mm}}{\makecell[c]{45\%}}&\multicolumn{12}{p{18mm}}{\makecell[c]{39.17\%}}&\multicolumn{12}{p{18mm}}{\makecell[c]{10.83\%}} \\ \bottomrule[1.0pt]
    \end{tabular}
    \vspace{0.1cm}
    \caption{Participant profiles.}
    \label{tab:participantprofiles}
    \vspace{-0.5cm}
\end{table*}

{\small
\begin{enumerate}
    \item The first page introduced the experiment and asked profiling questions such as age, level of studies and self-assessed technological proficiency. We also asked their given/first name to improve \system's interaction capabilities and their phone number to start the call.
    \item The participants were then redirected to a page with all of the necessary disclosure information in order to comply with GDPR and the relevant ethical considerations. The participants could only proceed if they explicitly agreed with the conditions.
    \item Next the instructions page is presented. After watching the instructions video in full and scrolling through all of the instructions, the participants could click the `start calls' button, which redirected to the next page with all of the necessary information about Innovatech and promptly instructed \system to call the participant.
    \item The participants then land in a page with all the provided information about Innovatech while receiving three calls. Only after fully completing the calls were the participants given the opportunity to advance.
    \item A page with questions about the social capabilities of the tool is presented. At this stage participants were not yet told that the calls are from a bot.
    \item After answering the previous questions, we disclose that the calls were performed by \system and ask the participants additional questions about its realism.
    \item In the last page, we invite the participants to give open feedback about the experiment. Although it was not graded, the feedback was useful for us to understand the general experience and perceived recommendations.
\end{enumerate}
}

The questionnaires and \system were interconnected via a REST API, which allowed \system to call the participants and the web application to know when the calls finished. All data was collected in a SQLite database.

\begin{table*}[t]
    \centering
    \begin{minipage}{.475\linewidth}
        \centering
        \footnotesize
        \begin{tabular}{>{\RaggedRight\arraybackslash}p{5cm} | >{\RaggedLeft\arraybackslash}p{.9cm} l p{.1cm}} 
        \hline
        \multicolumn{4}{>{\centering\arraybackslash}p{8cm}}{\textbf{Q1. }How would you describe your initial impression of the caller?}\\
        \hline
        \multicolumn{1}{c}{Answer} & \multicolumn{3}{c}{Frequency}\\
        \hline
        \textbf{1. }Very negative & 2.08\% & \ChartRed{.0208} & \\
        \textbf{2. }Somewhat negative & 7.5\% & \ChartOrange{.075} & \\
        \textbf{3. }Neutral & 27.5\% & \ChartYellow{.275} & \\
        \textbf{4. }Somewhat positive & 34.17\% & \ChartLGreen{.3417} & \\
        \textbf{5. }Very positive & 28.75\% & \ChartGreen{.2875} & \\
        \hline
        \multicolumn{4}{>{\centering\arraybackslash}p{8cm}}{\textbf{Q2. }Did anything about the caller's voice or manner of speaking stand out to you?}\\
        \hline
        \multicolumn{1}{c}{Answer} & \multicolumn{3}{c}{Frequency}\\
        \hline
        \textbf{1. }Very unusual or striking & 6.25\% & \ChartRed{.0625} & \\
        \textbf{2. }Somewhat unusual & 23.75\% & \ChartOrange{.2375} & \\
        \textbf{3. }Neutral, nothing particular & 27.5\% & \ChartYellow{.275} & \\
        \textbf{4. }Pleasant and engaging & 23.75\% & \ChartLGreen{.2375} & \\
        \textbf{5. }Very professional and clear & 18.75\% & \ChartGreen{.1875} & \\
        \hline
        \multicolumn{4}{>{\centering\arraybackslash}p{8cm}}{\textbf{Q3. }How natural did the conversation with the caller feel?}\\
        \hline
        \multicolumn{1}{c}{Answer} & \multicolumn{3}{c}{Frequency}\\
        \hline
        \textbf{1. }Extremely unnatural & 7.92\% & \ChartRed{.0792} & \\
        \textbf{2. }Somewhat unnatural & 27.5\% & \ChartOrange{.275} & \\
        \textbf{3. }Neutral & 14.17\% & \ChartYellow{.1417} & \\
        \textbf{4. }Mostly natural & 41.25\% & \ChartLGreen{.4125} & \\
        \textbf{5. }Completely natural & 9.17\% & \ChartGreen{.0917} & \\
        \hline
        \multicolumn{4}{>{\centering\arraybackslash}p{8cm}}{\textbf{Q4. }Were there any moments in the conversation that felt scripted or unnatural?}\\
        \hline
        \multicolumn{1}{c}{Answer} & \multicolumn{3}{c}{Frequency}\\
        \hline
        \textbf{1. }The entire conversation felt scripted & 13.33\% & \ChartRed{.1333} & \\
        \textbf{2. }Many parts felt scripted & 29.17\% & \ChartOrange{.2917} & \\
        \textbf{3. }Some scripted moments & 40.42\% & \ChartYellow{.4042} & \\
        \textbf{4. }Rarely scripted & 11.67\% & \ChartLGreen{.1167} & \\
        \textbf{5. }Completely spontaneous & 5.42\% & \ChartGreen{.0542} & \\
        \hline
        \multicolumn{4}{>{\centering\arraybackslash}p{8cm}}{\textbf{Q5. }Did the caller come across as credible and trustworthy?}\\
        \hline
        \multicolumn{1}{c}{Answer} & \multicolumn{3}{c}{Frequency}\\
        \hline
        \textbf{1. }Not credible at all & 10.83\% & \ChartRed{.1083} & \\
        \textbf{2. }Somewhat credible & 20.83\% & \ChartOrange{.2083} & \\
        \textbf{3. }Neutral & 22.08\% & \ChartYellow{.2208} & \\
        \textbf{4. }Mostly credible & 32.5\% & \ChartLGreen{.325} & \\
        \textbf{5. }Highly credible & 13.75\% & \ChartGreen{.1375} & \\
        \hline
        \end{tabular}%
    \end{minipage}%
    \hfill
    \begin{minipage}{.475\linewidth}
        \centering
        \footnotesize
        \begin{tabular}{>{\RaggedRight\arraybackslash}p{5cm} | >{\RaggedLeft\arraybackslash}p{.9cm} l p{.1cm}} 
        \hline
        \multicolumn{4}{>{\centering\arraybackslash}p{8cm}}{\textbf{Q6. }Did you feel comfortable sharing information with the caller?}\\
        \hline
        \multicolumn{1}{c}{Answer} & \multicolumn{3}{c}{Frequency}\\
        \hline
        \textbf{1. }Very uncomfortable & 12.5\% & \ChartRed{.125} & \\
        \textbf{2. }Somewhat uncomfortable & 28.75\% & \ChartOrange{.2875} & \\
        \textbf{3. }Neutral & 17.08\% & \ChartYellow{.1708} & \\
        \textbf{4. }Mostly comfortable & 26.67\% & \ChartLGreen{.2667} & \\
        \textbf{5. }Completely comfortable & 15.0\% & \ChartGreen{.15} & \\
        \hline
        \multicolumn{4}{>{\centering\arraybackslash}p{8cm}}{\textbf{Q7. }What motivated you to share (or not share) information with the caller?}\\
        \hline
        \multicolumn{1}{c}{Answer} & \multicolumn{3}{c}{Frequency}\\
        \hline
        \textbf{1. }Strong distrust or discomfort & 8.33\% & \ChartRed{.0833} & \\
        \textbf{2. }Mild distrust or discomfort & 22.92\% & \ChartOrange{.2292} & \\
        \textbf{3. }Neutral, no specific motivation & 25.83\% & \ChartYellow{.2583} & \\
        \textbf{4. }Felt somewhat obliged or interested & 35.42\% & \ChartLGreen{.3542} & \\
        \textbf{5. }Strong trust or interest in sharing & 7.5\% & \ChartGreen{.075} & \\
        \hline
        \multicolumn{4}{>{\centering\arraybackslash}p{8cm}}{\textbf{Q8. }What did you think was the caller's intent or goal during the conversation?}\\
        \hline
        \multicolumn{1}{c}{Answer} & \multicolumn{3}{c}{Frequency}\\
        \hline
        \textbf{1. }Purely sales-oriented or persuasive & 4.58\% & \ChartRed{.0458} & \\
        \textbf{2. }Primarily informational with some sales intent & 15.42\% & \ChartOrange{.1542} & \\
        \textbf{3. }Balanced between information and sales & 38.33\% & \ChartYellow{.3833} & \\
        \textbf{4. }Mostly informational & 31.25\% & \ChartLGreen{.3125} & \\
        \textbf{5. }Completely informational or consultative & 10.42\% & \ChartGreen{.1042} & \\
        \hline
        \multicolumn{4}{>{\centering\arraybackslash}p{8cm}}{\textbf{Q9. }Did the conversation elicit any emotional response from you, such as frustration, curiosity, or confidence?}\\
        \hline
        \multicolumn{1}{c}{Answer} & \multicolumn{3}{c}{Frequency}\\
        \hline
        \textbf{1. }Strong negative emotions (e.g., frustration, anger) & 2.92\% & \ChartRed{.0292} & \\
        \textbf{2. }Mild negative emotions & 17.92\% & \ChartOrange{.1792} & \\
        \textbf{3. }Neutral, no emotional response & 48.75\% & \ChartYellow{.4875} & \\
        \textbf{4. }Mild positive emotions (e.g., curiosity, interest) & 26.25\% & \ChartLGreen{.2625} & \\
        \textbf{5. }Strong positive emotions (e.g., confidence, satisfaction) & 4.17\% & \ChartGreen{.0417} & \\
        \hline
        \multicolumn{4}{>{\centering\arraybackslash}p{8cm}}{\textbf{Q10. }How did the caller's approach influence your emotional response?}\\
        \hline
        \multicolumn{1}{c}{Answer} & \multicolumn{3}{c}{Frequency}\\
        \hline
        \textbf{1. }Led to strong negative emotions & 1.67\% & \ChartRed{.0167} & \\
        \textbf{2. }Caused some negative feelings & 16.67\% & \ChartOrange{.1667} & \\
        \textbf{3. }No significant influence on emotions & 51.67\% & \ChartYellow{.5167} & \\
        \textbf{4. }Contributed to positive feelings & 25.42\% & \ChartLGreen{.2542} & \\
        \textbf{5. }Significantly boosted positive emotions & 4.58\% & \ChartGreen{.0458} & \\
        \hline 
        \end{tabular}%
    \end{minipage}%
\vspace{0.1cm}
\caption{\label{tab:table_eff_main} \textbf{Form 1: Effectiveness and trustworthiness questionnaire:} Complete set of questions and answers to the first part of the feedback section of the experiment; the participants were not aware that they were talking with an AI-powered vishing system.}
\end{table*}

\begin{table*}[t]
    \centering
    \begin{minipage}{.475\linewidth}
        \centering
        \footnotesize
        \begin{tabular}{>{\RaggedRight\arraybackslash}p{5cm} | >{\RaggedLeft\arraybackslash}p{.9cm} l p{.1cm}}
        \hline
        \multicolumn{4}{>{\centering\arraybackslash}p{8cm}}{\textbf{Q1. }Knowing now that the caller was an AI, how realistic do you find the voice quality?}\\
        \hline
        \multicolumn{1}{c}{Answer} & \multicolumn{3}{c}{Frequency}\\
        \hline
        \textbf{1. }Completely artificial & 10.42\% & \ChartRed{.1042} & \\
        \textbf{2. }Mostly artificial & 24.17\% & \ChartOrange{.2417} & \\
        \textbf{3. }Neutral & 13.75\% & \ChartYellow{.1375} & \\
        \textbf{4. }Mostly realistic & 40.42\% & \ChartLGreen{.4042} & \\
        \textbf{5. }Completely realistic & 11.25\% & \ChartGreen{.1125} & \\
        \hline
        \multicolumn{4}{>{\centering\arraybackslash}p{8cm}}{\textbf{Q2. }How would you rate the believability of the voice?}\\
        \hline
        \multicolumn{1}{c}{Answer} & \multicolumn{3}{c}{Frequency}\\
        \hline
        \textbf{1. }Completely artificial &  9.17\% & \ChartRed{.0917} & \\
        \textbf{2. }Mostly artificial & 25.42\% & \ChartOrange{.2542} & \\
        \textbf{3. }Neutral & 18.75\% & \ChartYellow{.1875} & \\
        \textbf{4. }Mostly human-like & 40.42\% & \ChartLGreen{.4042} & \\
        \textbf{5. }Indistinguishable from a human &  6.25\% & \ChartGreen{.0625} & \\
        \hline
        \multicolumn{4}{>{\centering\arraybackslash}p{8cm}}{\textbf{Q3. }Did you notice any lags or delays in the conversation?}\\
        \hline
        \multicolumn{1}{c}{Answer} & \multicolumn{3}{c}{Frequency}\\
        \hline
        \textbf{1. }Frequent lags and delays & 10.42\% & \ChartRed{.1042} & \\
        \textbf{2. }Occasional lags and delays & 42.92\% & \ChartOrange{.4292} & \\
        \textbf{3. }Rare lags and delays & 21.25\% & \ChartYellow{.2125} & \\
        \textbf{4. }Very rare lags and delays & 15.83\% & \ChartLGreen{.1583} & \\
        \textbf{5. }No lags or delays &  9.58\% & \ChartGreen{ .0958} & \\
        \hline
        \multicolumn{4}{>{\centering\arraybackslash}p{8cm}}{\textbf{Q4. }How effectively did the AI respond to your questions and comments?}\\
        \hline
        \multicolumn{1}{c}{Answer} & \multicolumn{3}{c}{Frequency}\\
        \hline
        \textbf{1. }Very ineffectively &  2.92\% & \ChartRed{ .0292} & \\
        \textbf{2. }Somewhat ineffectively & 11.67\% & \ChartOrange{.1167} & \\
        \textbf{3. }Neutral & 14.17\% & \ChartYellow{.1417} & \\
        \textbf{4. }Mostly effectively & 53.33\% & \ChartLGreen{.5333} & \\
        \textbf{5. }Very effectively & 17.92\% & \ChartGreen{.1792} & \\
        \hline 
        \multicolumn{4}{>{\centering\arraybackslash}p{8cm}}{\textbf{Q5. }Were there any moments in the conversation where the AI's responses seemed inappropriate or out of context?}\\
        \hline
        \multicolumn{1}{c}{Answer} & \multicolumn{3}{c}{Frequency}\\
        \hline
        \textbf{1. }Frequently inappropriate &  4.58\% & \ChartRed{ .0458} & \\
        \textbf{2. }Occasionally inappropriate & 16.67\% & \ChartOrange{.1667} & \\
        \textbf{3. }Rarely inappropriate & 19.17\% & \ChartYellow{.1917} & \\
        \textbf{4. }Very rarely inappropriate & 24.17\% & \ChartLGreen{.2417} & \\
        \textbf{5. }Always appropriate and in context & 35.42\% & \ChartGreen{.3542} & \\
        \hline\end{tabular}%
    \end{minipage}%
    \hfill
    \begin{minipage}{.475\linewidth}
        \centering
        \footnotesize
        \begin{tabular}{>{\RaggedRight\arraybackslash}p{5cm} | >{\RaggedLeft\arraybackslash}p{.9cm} l p{.1cm}}
        \hline
        \multicolumn{4}{>{\centering\arraybackslash}p{8cm}}{\textbf{Q6. }Did you encounter any technical issues during the call, such as voice breaking, echoes, or other anomalies?}\\
        \hline
        \multicolumn{1}{c}{Answer} & \multicolumn{3}{c}{Frequency}\\
        \hline
        \textbf{1. }Multiple technical issues &  2.50\% & \ChartRed{.0250} & \\
        \textbf{2. }Some technical issues &  5.00\% & \ChartOrange{.0500} & \\
        \textbf{3. }Few technical issues &  9.58\% & \ChartYellow{.0958} & \\
        \textbf{4. }Very few technical issues & 21.25\% & \ChartLGreen{.2125} & \\
        \textbf{5. }No technical issues & 61.67\% & \ChartGreen{.6167} & \\
        \hline
        \multicolumn{4}{>{\centering\arraybackslash}p{8cm}}{\textbf{Q7. }How does your experience with the AI caller compare to typical phone conversations you have with real people?}\\
        \hline
        \multicolumn{1}{c}{Answer} & \multicolumn{3}{c}{Frequency}\\ 
        \hline
        \textbf{1. }Significantly worse than with real people & 12.08\% & \ChartRed{.1208} & \\
        \textbf{2. }Somewhat worse than with real people & 25.00\% & \ChartOrange{.25} & \\
        \textbf{3. }Comparable to real people & 42.50\% & \ChartYellow{.4250} & \\
        \textbf{4. }Better than with most real people & 15.42\% & \ChartLGreen{.1542} & \\
        \textbf{5. }Significantly better than with real people &  5.00\% & \ChartGreen{.05} & \\
        \hline
        \multicolumn{4}{>{\centering\arraybackslash}p{8cm}}{\textbf{Q8. }At what point, if ever, did you start to suspect the caller might be an AI?}\\
        \hline
        \multicolumn{1}{c}{Answer} & \multicolumn{3}{c}{Frequency}\\
        \hline
        \textbf{1. }Immediately from the beginning & 56.25\% & \ChartRed{.5625} & \\
        \textbf{2. }After a few exchanges & 25.83\% & \ChartOrange{.2583} & \\
        \textbf{3. }Midway through the conversation &  7.92\% & \ChartYellow{.0792} & \\
        \textbf{4. }Towards the end of the conversation &  6.25\% & \ChartLGreen{.0625} & \\
        \textbf{5. }Never suspected it was an AI &  3.75\% & \ChartGreen{.0375} & \\
        \hline
        \multicolumn{4}{>{\centering\arraybackslash}p{8cm}}{\textbf{Q9. }What specific elements of the conversation led you to this suspicion?}\\
        \hline
        \multicolumn{1}{c}{Answer} & \multicolumn{3}{c}{Frequency}\\
        \hline
        \textbf{1. }Unnatural voice or speech patterns & 27.50\% & \ChartRed{.2750} & \\
        \textbf{2. }Lack of emotional response or empathy & 24.17\% & \ChartOrange{.2417} & \\
        \textbf{3. }Repetitive or irrelevant responses & 12.08\% & \ChartYellow{.1208} & \\
        \textbf{4. }Delayed or slow responses & 20.83\% & \ChartLGreen{.2083} & \\
        \textbf{5. }No specific elements, just a general feeling & 15.42\% & \ChartGreen{.1542} & \\
        \hline \end{tabular}%
    \end{minipage}%
\vspace{0.1cm}
\caption{\label{tab:table_real_main} \textbf{Form 2: Perceived realism questionnaire:} Complete set of questions and answers to the second part of the feedback section of the experiment; after the participants were made aware they were talking with an AI-powered vishing system.}
\end{table*}

\end{document}